\documentclass[aps, onecolumn, pre, floatfix, superscriptaddress, 10pt]{revtex4-2}
\usepackage{graphicx} 
\usepackage{subcaption} 
\usepackage{pgfplots,float,pgfplotstable} 
\pgfplotsset{compat=1.15}

\usepackage{csvsimple} 
\usepackage{bm}
\usepackage{bbold}
\usepackage{physics,amsmath,amssymb,amsfonts,amsthm,mathtools}
\usepackage{ulem}   
\usepackage{comment}
\usepackage{bbm}

\usepackage{hyperref}
\hypersetup{
	colorlinks,    
	linkcolor=blue,     
	citecolor=blue,   
	urlcolor=blue    
}	

\usepackage{listings}
\newfloat{Figure}{htbp}{figs}
\graphicspath{{./figures/}}

\pgfplotstableset{
	my num iterating settings/.style={%
		/pgfplots/table/display columns/#1/.style={
			postproc cell content/.style={
				@cell content/.add={\small}{}
			}
		}
	}
}

\usepackage{array, multirow, makecell} 
\newcolumntype{R}[1]{>{\raggedleft\arraybackslash }b{#1}}
\newcolumntype{L}[1]{>{\raggedright\arraybackslash }b{#1}}
\newcolumntype{C}[1]{>{\centering\arraybackslash }b{#1}}

\definecolor{darkgreen}{rgb}{0,0.5,0.5}

\def\rev#1{\textcolor{black}{#1}}

\begin{document}

    \title{First passage time in space-dependent stochastic resetting}

	\author{Johannes Aspman}
	\affiliation{Department of Computer Science,
		Czech Technical University in Prague, Czech Republic}

        \author{Daniel Mastropietro}
        \affiliation{Department of Computer Science,
		Czech Technical University in Prague, Czech Republic}
        \affiliation{Universit\'e de Toulouse INP, 31071 Toulouse, France}
        \affiliation{CNRS-IRIT, 31071 Toulouse, France}
	
	\author{Jakub Mare\v{c}ek}
	\affiliation{Department of Computer Science,
		Czech Technical University in Prague, Czech Republic}

	\date{\today}

	\begin{abstract}
		
		We consider the mean first passage time (MFPT) through a target of interest for a diffusive particle of Langevin type, with the added condition that the particle is reset to its original position with some rate $r$. We study both smooth and non-smooth, non-convex potentials, focusing on the case where the reset rate depends on the space coordinate. For quadratic and piecewise-quadratic potentials, we show that the benefits of resetting depend on the ratio between drift and noise, and become more important as the drift potential becomes smaller compared to the noise. When the target is a local optimum of the potential, we further show that it is beneficial to use a space-dependent resetting where the reset rate is lower when the particle is closer to the target.
	\end{abstract}
	
	\maketitle
	
	\section{Introduction}
Searching for a given target is something we do every day, and it is a ubiqituous problem in modern technology and research. For example, enzymes searching for a target site in a DNA strand \cite{eliazar2007searching}, neural network training algorithms searching for an optimum \cite{lecun2015deep}, or simply when we are looking for our keys in the morning. Resetting strategies for searching are built on the simple idea that after performing the search for some amount of time a natural step might be to return to a given point and start your search again. 

A natural question is to ask whether resetting provably speeds up the search. 
In their seminal papers \cite{evans2011diffusion,evans2011diffusionA,evans2014diffusion}, Evans and Majumdar answer the question in a model, where a diffusion process has a non-zero probability rate of resetting to a given fixed point. Similar models have also been proposed by other authors  \cite{benichou2007intermittent,gelenbe2010search,janson2012hitting}, and many extensions of Evans and Majumdars approach have been studied recently \cite{pal2017first,roldan2017path,evans2020stochastic,ahmad2019first,pal2019landau,ray2019peclet,ray2020diffusion, bonomo2021first,ahmad2022first}, largely within statistical mechanics. 
Resetting has also been studied in mathematical optimisation for many years as a way of speeding up algorithms \cite{luby1993optimal,alt1996method}. In AI and computer science, there is a long history of engineering local-search algorithms with resets \cite{kautz2002dynamic,tong2008random,Shylo2018,yu2019parallel, loshchilov2022sgdr, bae2024stochastic}. 
In particular, much of the literature  \cite{gagliolo2007learning, wedge2008heavy,roulet2017sharpness,lorenz2021restart,renegar2022simple,starkov2023universal} \rev{focuses on the case where the drift potential is convex and smooth.} For example,  \cite{roulet2017sharpness, renegar2022simple} studies various first-order methods and show that resetting can be used to accelerate these. \cite{lorenz2021restart} discusses Luby's \cite{luby1993optimal} reset strategy in a continuous setting and \cite{starkov2023universal} derives universal performance bounds on general resetting strategies for stochastic processes.
The non-smooth non-convex setting \cite{davis2020stochastic,berner2021modern}, while clearly relevant in AI and machine learning \rev{due to the ubiquitous use of neural network modeling via non-smooth neuron activation functions},  
has not been given much attention in the resetting literature.
The noteworthy exception is Ahmad et al.~\cite{ahmad2022first}, who study piecewise-linear potentials as well as smooth, but non-convex, potentials. 

In this paper, we extend the work of Evans and Majumdar \cite{evans2011diffusion,evans2011diffusionA,evans2014diffusion, evans2020stochastic}, as well as the work of Ahmad et al. \cite{ahmad2022first} to non-smooth, non-convex potentials and reset rates conditional on the gradient of the objective function, whose global optimum is sought. In the case where the optimum is known, it may seem intuitive to reset less often, once a threshold close to the optimum is reached, for instance.
Note that this is an important special case, considering that the empirical risk (the error of a model's prediction of a target of interest) is bounded from below by 0 in most machine learning applications, but it is also bounded from above by 0 in the overparameterized regime \cite{lecun2015deep}, typical of many modern models where the number of parameters is far larger than the number of training cases. In the more general case where the optimum is unknown, an indication of such a situation is partially given by a threshold on the gradient of the objective function: assuming the optimum occurs in a smooth region of the parameter space, a ``small'' value of the gradient norm is an indication of being close to a local optimum. This is the case we focus on in this work.
Specifically, we consider a piecewise quadratic potential for the diffusion process, and we introduce a piecewise-constant reset rate that depends on the gradient of the objective function. 
We show that the mean first passage time (MFPT) can be reduced when one considers the piecewise-constant reset rate, conditional on the gradient norm of the objective function, rather than any single reset rate, or no reset at all. For simplicity, we focus mainly on the one-dimensional case, but provide a brief comparison with the higher-dimensional situation.
We also present numerical results in both the one- and two-dimensional cases that support the theoretical findings.

\section{Problem statement and background}
\label{sec:problem}
As mentioned in the introduction, we will not consider a pure diffusion, but rather diffusion with drift, or diffusion in a potential landscape. In other words, we are considering a process $x(t) \in \mathbb{R}^d$ satisfying
\begin{equation}
\label{eq:diffusion_equation}
    \dot x= \mu F(x)+\eta(t),
\end{equation}
where $\dot x$ denotes the temporal derivative of $x(t)$ and $F(x)$ is given by the negative spatial gradient of a potential $V(x) : \mathbb{R}^d \to \mathbb{R}$,
\begin{equation}
F(x)= -\nabla V(x)
\end{equation}
or a suitable generalisation thereof in the non-smooth setting (e.g. the subgradient). In general, parameter $\mu$ suggests the mobility of the particle, although we consider only $\mu=1$ in the rest of the paper. The noise $\eta(t) \coloneqq (\eta_i(t))_{i=1, \dots, d}$ is assumed to be a white Gaussian vector with the property:
\begin{equation*}
    \langle \eta_i(t)\rangle=0,\quad \langle\eta_i(t)\eta_j(t')\rangle=2D\delta_{ij}\delta(t-t'), \qquad \text{for } i,j = 1, \dots, d,
\end{equation*}
where $D>0$ is called the diffusion coefficient, $\delta_{ij}$ is the Kronecker delta, and $\delta(t)$ the Dirac delta function \cite{pal2015diffusion}.

We will focus our analysis on the mean first passage time (MFPT) $T(x)$ for the particle to reach a given target $L$ from the point $x$. It is well-known \cite{redner2001guide}, that the MFPT is infinite for a pure diffusive particle (or random search), while \cite{evans2011diffusion} showed that in the presence of resetting, it becomes finite.  The standard starting point for MFPT analysis is the backward Chapman-Kolmogorov equation, which governs the probability $Q(x,t)$ that the particle has not reached the desired target up to time $t$, starting from position $x$. For $x\in\mathbb{R}$, the equation reads 
\begin{equation}\label{eq:chapman_kolmogorov}
    \frac{\partial Q(x,t)}{\partial t}=D\frac{\partial^2 Q(x,t)}{\partial x^2}- V'(x)\frac{\partial  Q(x,t)}{\partial x}-r(x)Q(x,t)+r(x)Q(x_0,t), 
\end{equation}
with boundary conditions $Q(L,t)=0$, $Q(x,0)=1$, where $x_0$ is the initial point, as well as the point where the process restarts. Noting that $-\tfrac{\partial Q(x,t)}{\partial t}\mathrm dt$ measures the probability that the particle hits the target in the time interval $[t, t+\mathrm dt)$, we find the MFPT by integrating \eqref{eq:chapman_kolmogorov} and using that
\begin{equation*}
    T(x)=-\int_0^\infty t\frac{\partial Q(x,t)}{\partial t}\mathrm dt=\int_0^\infty Q(x,t)\mathrm dt,
\end{equation*}
to obtain the equation
\begin{equation}\label{eq:gen_MFPT}
    -1=D\frac{\partial^2T(x)}{\partial x^2}- V'(x)\frac{\partial T(x)}{\partial x}-r(x)T(x)+r(x)T(x_0),
\end{equation}
for the MFPT, with boundary conditions $T(L)=0$ and $T(x)$ remains finite as $x\to\infty$ \cite{evans2011diffusionA, evans2020stochastic}. 

Ahmad et al. \cite{ahmad2019first} study the situation where the potential is a polynomial, $V(x)=kx^n$ with $k,n>0$, and where the reset rate is constant. (See also \cite{ray2019peclet}, for the linear case, and \cite{ray2020diffusion} for the case of a logarithmic potential, $V(x)=V_0\log(|x|)$. A piecewise linear potential was studied in \cite{ahmad2022first}.) We will expand on this in two ways. Firstly, instead of a constant reset rate, we will consider having a piecewise-constant reset rate, i.e., where the reset rate is allowed to change when we move around in space. Secondly, rather than smooth polynomial potentials \cite{ahmad2019first} or piecewise linear potentials \cite{ahmad2022first}, we will consider piecewise polynomial potentials, in general, and piecewise quadratic potentials, in particular. The motivation for studying the piecewise quadratic potential comes from modern machine learning applications, where this is the predominant structure of the \rev{objective function} \cite{berner2021modern}. The inspiration for studying the effect of having a piecewise-constant reset rate comes from different types of informed reset strategies, such as gradient-informed reset \cite{mastropietro2026ieeequantum} which have been shown to be beneficial in many situations. 

\begin{figure}[tp]
\centering
    \begin{subfigure}{.48\linewidth}
    \includegraphics[width=\linewidth]{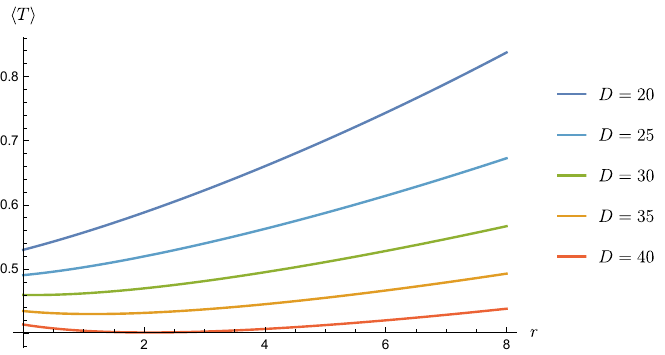}
    \caption{}
    \label{subfig:smooth_constant_a}
    \end{subfigure}
    \hfill
    \begin{subfigure}{.48\linewidth}
    \includegraphics[width=.7\linewidth]{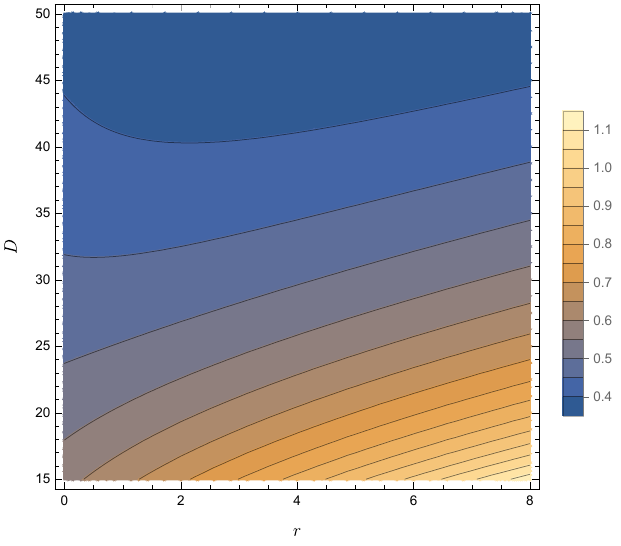}
    \caption{}
    \label{subfig:smooth_constant_b}
    \end{subfigure}
\caption{(1d smooth potential / constant reset / varying $D$) MFPT, $\langle{T}\rangle \coloneqq T_0$, of reaching a target $x=L$ for the quadratic potential $V(x)=x^2$, computed using Eq.~\eqref{eq:smooth_sol_constR}.
Here we have fixed $L=0.01$ and $x_0=4$. \newline
(a) MFPT as a function of the constant reset rate $r$, for values of $D \in \{20, 25, 30, 35, 40\}$.
(b) Contour plot of MFPT as a function of $r$ and $D$.
All numerical evaluations were done using Mathematica \cite{Mathematica}.
}
\label{fig:smooth_constant}
    \end{figure}

Let us review some of the previous results from \cite{ahmad2019first}. For brevity, and as previously mentioned, we assume that the particle will reset from the initial position $x=x_0$. In general, Equation \eqref{eq:gen_MFPT} is very difficult to solve, but when $r(x)$ and $V(x)$ take some simple forms, we can solve it analytically. 
When we have a constant reset rate $r(x)=r$ and polynomial potential $V(x)=kx^n$ we get the equation
\begin{equation}\label{eq:quad_eq}
    -1=D\frac{\partial^2 T(x)}{\partial x^2}-knx^{n-1}\frac{\partial T(x)}{\partial x}-rT(x)+rT_0,
\end{equation}
where we introduced the notation
$T_0 \coloneqq T(x_0)$.
Analytical solutions do not exist for $n>3$, so we focus on the case of $n=2$, which was also studied in \cite{ahmad2019first}. The general solutions are now
\begin{equation}\label{eq:gen_sol}
        T(x)=AH_{-\tfrac{r}{2k}}\left(\sqrt{\tfrac{k}{D}}x\right)+B 
        F_1\left[\tfrac{r}{4k},\tfrac12;\tfrac{k}{D}x^2\right]+\tfrac1r+T_0,
\end{equation}
where $H_n(x)$ are the Hermite polynomials, ${}_1F_1[a,b;x]$ is the Kummer confluent hypergeometric function and $A$, $B$ are scalars to be determined. The boundary conditions constrain the solution \eqref{eq:gen_sol} to be finite for $x\to\infty$ and $T(L)=0$, for some target $x=L$. These conditions determine $A$ and $B$, and, after some algebraic manipulations, we find the solution of the MFPT for the smooth quadratic potential using a constant reset rate:
\begin{equation}\label{eq:smooth_sol_constR}
    T_0=\frac{1}{r}\left(\frac{H_{-\tfrac{r}{2k}}\left(\sqrt{\tfrac{k}{D}}L\right)}{H_{-\tfrac{r}{2k}}\left(\sqrt{\tfrac{k}{D}}x_0\right)}-1\right).
\end{equation}
This agrees with \cite{ahmad2019first}. When the potential is weak, i.e., for low signal-to-noise ratio represented by a small initial drift compared to the noise level, one can find that there is an optimal non-zero reset rate, cf. \cite{ahmad2019first}, which means that it is beneficial to use resetting for this scenario. As can be seen in Fig.~\ref{fig:smooth_constant} on page \pageref{fig:smooth_constant}, there are non-zero local minima of the MFPT for a large-enough value of $D$. In \cite{ahmad2019first}, the phase transition is found numerically and expressed in terms of the dimensionless variable $K \coloneqq |V'(x_0)| / \sqrt{4D} = k x_0/\sqrt{D}$, whose critical value is $K_c\approx 0.7393$. That is, an optimal reset rate exists only when $K < K_c$.
This is verified in Fig.~\ref{fig:smooth_constant}, which plots the MFPT, $T_0$ in \eqref{eq:smooth_sol_constR}, for selected values of $D$ ranging from $20$ to $40$ when $k=1$, $x_0=4$, $L=0.01$: the minimum disappears when $D$ goes below $30$, which is consistent with the value of $K_c$ above since, for $k=1$ and $x_0=4$, the critical value corresponds to $D \approx 29.3$.

We end the section by noting that here, as in later sections, it would also make sense to consider only the case $L=0$, as this is the global minimum, with the advantage that the expression for $T_0$ is further simplified. However, to keep expressions general, we choose $L$ small compared to $x_0$, but non-zero.

\section{Theoretical results}
\label{sec:results_theory}
Let us now turn to our analysis. We start by considering the smooth quadratic potential as above, but with a piecewise-constant reset rate. After this we consider the situations where we have a non-smooth, piecewise quadratic potential both with a constant and with a piecewise-constant reset rate.

\subsection{Smooth quadratic potential: piecewise-constant reset rate}
As discussed in the introduction, it is natural to consider having a reset rate that depends on the space time coordinate $x$. Some simple situations have been studied in the literature; see, for example, \cite{evans2011diffusionA, roldan2017path}. We will consider a version of this in which $r(x)$ changes between two different constant rates depending on whether the value of some function $f(x)$ is within a threshold. In particular, we consider
\begin{equation}\label{eq:2resets}
    r(x)=\begin{cases}
        r_1,\quad &|f(x)|>\beta,\\
        r_2,\quad &|f(x)|\leq \beta,
    \end{cases}
\end{equation}
for some function $f(x)$ and some constant $\beta>0$. We will restrict ourselves 
to the case where the condition is governed by the magnitude of the gradient (such as $f(x)=V'(x)=2kx$ in the smooth case) or a suitable generalisation (to the non-smooth case). This means that we now have two copies of equation \eqref{eq:quad_eq}, one copy for the case with $r_1$ and one with $r_2$. The general solutions will also look the same in each window. As extra boundary conditions, we should now also demand that both $T(x)$ and $T'(x)$ be continuous on the boundary $|f(x)|=\beta$. The solution will depend on whether we set the target $x=L$ and the reset point $x=x_0$ below or above the threshold $|x|=\tfrac{\beta}{2k}$. We focus on the case where $L$ is close to the origin, so we assume $L<\tfrac{\beta}{2k}$, and where the resetting is further away, so that $x_0>\tfrac{\beta}{2k}$. After some algebraic manipulations, we find the solution of the MFPT for the smooth quadratic potential using a piecewise-constant reset rate:
\begin{equation}\label{eq:smooth_sol_2r}\footnotesize
    \begin{aligned}
        T_0=&\frac{1}{r_1r_2H_{-\tfrac{r_1}{2k}}\left(\sqrt{\tfrac{k}{D}}x_0\right)\left(2\sqrt{kD}H_{-1-\tfrac{r_2}{2k}}\left(\tfrac{{\beta}}{2\sqrt{kD}}\right){}_1F_1\left[\tfrac{r_2}{4k},\tfrac12;\tfrac{\beta^2}{4kD}\right]+\beta H_{-\tfrac{r_2}{2k}}\left(\tfrac{\beta}{2\sqrt{kD}}\right){}_1F_1\left[1+\tfrac{r_2}{4k},\tfrac32;\tfrac{\beta^2}{4kD}\right]\right)}\\
        &\times\Big[2r_1\sqrt{kD}H_{-1-\tfrac{r_1}{2k}}\left(\tfrac{\beta}{2\sqrt{kD}}\right)\left(H_{-\tfrac{r_2}{2k}}\left(\sqrt{\tfrac{k}{D}}L\right){}_1F_1\left[\tfrac{r_2}{4k},\tfrac12;\tfrac{\beta^2}{4kD}\right]\right)-H_{-\tfrac{r_2}{2k}}\left(\tfrac{\beta}{2\sqrt{kD}}\right){}_1F_1\left[\tfrac{r_2}{4k},\tfrac{1}{2};\tfrac{kL^2}{D}\right]\\
        &+r_2H_{-\tfrac{r_1}{2k}}\left(\tfrac{\beta}{2\sqrt{kD}}\right)\left(2\sqrt{kD}H_{-1-\tfrac{r_2}{2k}}\left(\tfrac{\beta}{2\sqrt{kD}}\right){}_1F_1\left[\tfrac{r_2}{4k},\tfrac{1}{2};\tfrac{kL^2}{D}\right]+\beta H_{-\tfrac{r_2}{2k}}\left(\sqrt{\tfrac{k}{D}}L\right){}_1F_1\left[1+\tfrac{r_2}{4k},\tfrac{3}{2};\tfrac{\beta^2}{4kD}\right]\right)\\
        &+H_{-\tfrac{r_1}{2k}}\left(\sqrt{\tfrac{k}{D}}x_0\right)\Big(2\sqrt{kD}H_{-1-\tfrac{r_2}{2k}}\left(\tfrac{\beta}{2\sqrt{kD}}\right)\left((r_1-r_2){}_1F_1\left[\tfrac{r_2}{4k},\tfrac12;\tfrac{kL^2}{D}\right]-r_1{}_1F_1\left[\tfrac{r_2}{4k},\tfrac12;\tfrac{\beta^2}{4kD}\right]\right)
        \\&+\beta\left((r_1-r_2)H_{-\tfrac{r_2}{2k}}\left(\sqrt{\tfrac{k}{D}}L\right)-r_1H_{-\tfrac{r_2}{2k}}\left(\tfrac{\beta}{2\sqrt{kD}}\right)\right){}_1F_1\left[1+\tfrac{r_2}{4k},\tfrac32;\tfrac{\beta^2}{4kD}\right]\Big)\Big].
    \end{aligned}
\end{equation}

\begin{figure}[tp]
\centering
    \begin{subfigure}{.48\linewidth}
    \includegraphics[width=\linewidth]{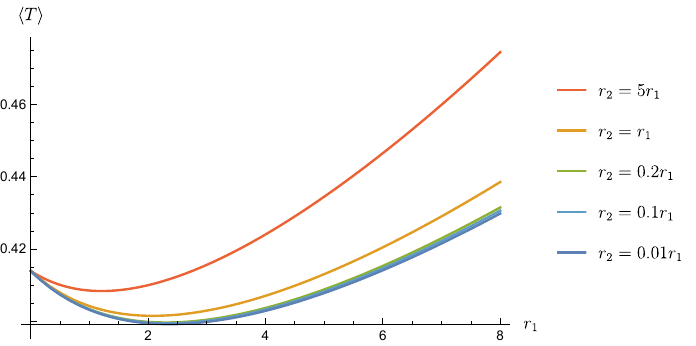}
    \caption{}
    \label{subfig:compare_smooth1_a}
    \end{subfigure}
    \hfill
    \begin{subfigure}{.48\linewidth}
    \includegraphics[width=\linewidth]{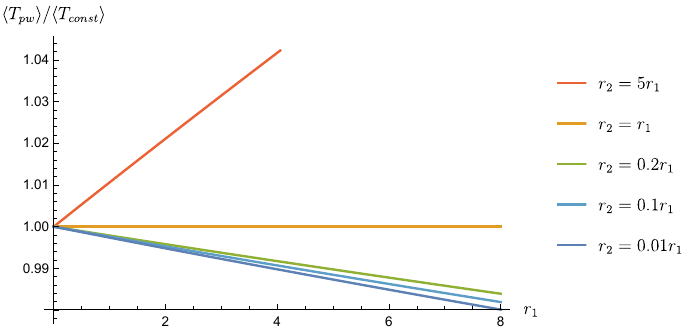}
    \caption{}
    \label{subfig:compare_smooth1_b}
    \end{subfigure}
\caption{(1d smooth potential / piecewise vs. constant reset / $D$ fixed)
Comparison of the MFPT, $\langle{T}\rangle \coloneqq T_0$ in Eq.~\eqref{eq:smooth_sol_2r}, between using a constant reset rate vs. the piecewise-constant reset rate defined in Eq.~\eqref{eq:2resets},
for the smooth quadratic potential $V(x) = x^2$. Here we have fixed $L=0.01$, $x_0=4$, $D=40$, $\beta=1$, and $r_2$ is set as a fixed multiple of $r_1$, i.e., the MFPT is plotted as a function of $r_1$ with $r_2=cr_1$, for values of $c \in \{ 0.01, 0.1, 0.2, 1, 5 \}$. 
\newline
(a) MFPT, labelled by $c$.
(b) Ratio between the piecewise-constant reset rate, $\langle{T_{pw}}\rangle$, and the constant reset rate, $\langle{T_{const}}\rangle$, labelled by $c$. 
All numerical evaluations were done using Mathematica \cite{Mathematica}.
}
\label{fig:compare_smooth1}
    \end{figure}

\begin{figure}[tp]
\centering
    \begin{subfigure}{.48\linewidth}
    \includegraphics[width=\linewidth]{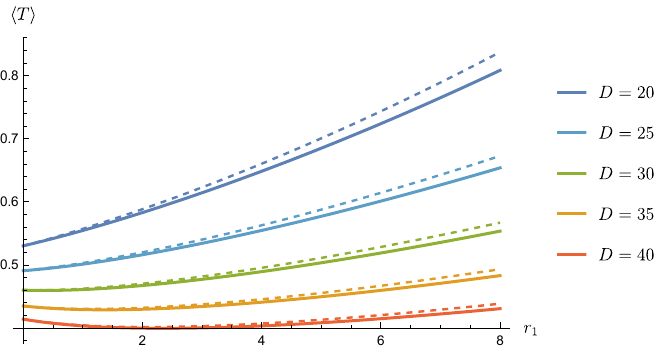}
    \caption{$r_2 = 0.1 r_1$}
    \label{subfig:compare_smooth2_a}
    \end{subfigure}
    \hfill
    \begin{subfigure}{.48\linewidth}
    \includegraphics[width=\linewidth]{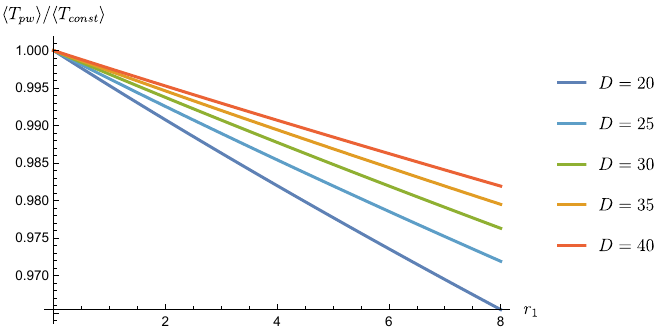}
    \caption{$r_2 = 0.1 r_1$}
    \label{subfig:compare_smooth2_b}
    \end{subfigure}
\caption{(1d smooth potential / piecewise vs. constant reset / varying $D$) Comparison of the MFPT, $\langle{T}\rangle \coloneqq T_0$ in Eq.~\eqref{eq:smooth_sol_2r}, between using a constant reset rate vs. the piecewise-constant reset rate defined in Eq.~\eqref{eq:2resets} with $r_2 = 0.1 r_1$, for the smooth quadratic potential $V(x) = x^2$, for values of $D \in \{20, 25, 30, 35, 40\}$. Here we have fixed $L=0.01$, $x_0=4$, $\beta=1$.
\newline
(a) MFPT for constant reset rate ($\langle{T_{const}}\rangle$, dashed) and piecewise-constant reset rate ($\langle{T_{pw}}\rangle$, solid), labelled by $D$.
(b) Ratio $\langle{T_{pw}}\rangle / \langle{T_{const}}\rangle$, labelled by $D$.
All numerical evaluations were done using Mathematica \cite{Mathematica}.
}
\label{fig:compare_smooth2}
    \end{figure}

An interesting analysis is to compare the results for our two cases, i.e., having one constant reset rate, as in Eq. \eqref{eq:smooth_sol_constR}, and having two different resets as in Eq. \eqref{eq:smooth_sol_2r}. This comparison is shown in Figs. \ref{fig:compare_smooth1} and \ref{fig:compare_smooth2}. There, we fix $r_2$ to be a multiple of $r_1$ for the case where we have two resets and plot $T_0$ as a function of $r_1$. We see that when $r_2>r_1$ it is always better to just keep the constant reset rate, while when $r_2<r_1$ it is better to switch the rate. This can be interpreted as saying that whenever we get closer to the target, it is better to not reset as often. 

\begin{figure*}[tp]
        \subfloat[$r_2=5r_1$]{%
            \includegraphics[width=.45\linewidth]{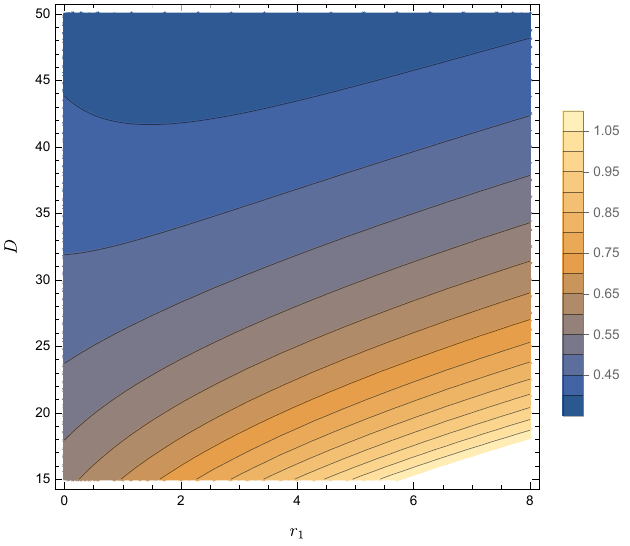}%
            \label{subfig:contour_a_smooth}%
        }\hfill
        \subfloat[$r_2=r_1$]{%
            \includegraphics[width=.45\linewidth]{figures/T0_smooth_constant_contour.pdf}%
            \label{subfig:contour_b_smooth}%
        }\\
        \subfloat[$r_2=0.1r_1$]{%
            \includegraphics[width=.45\linewidth]{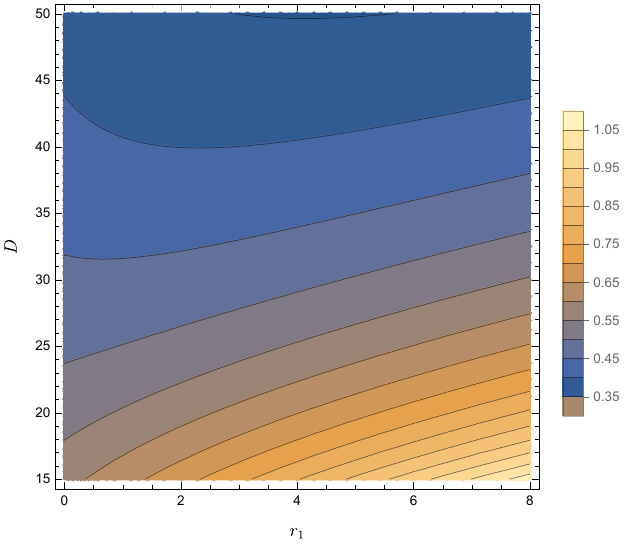}%
            \label{subfig:contour_c_smooth}%
        }\hfill
        \subfloat[$r_2=0.01r_1$]{%
            \includegraphics[width=.45\linewidth]{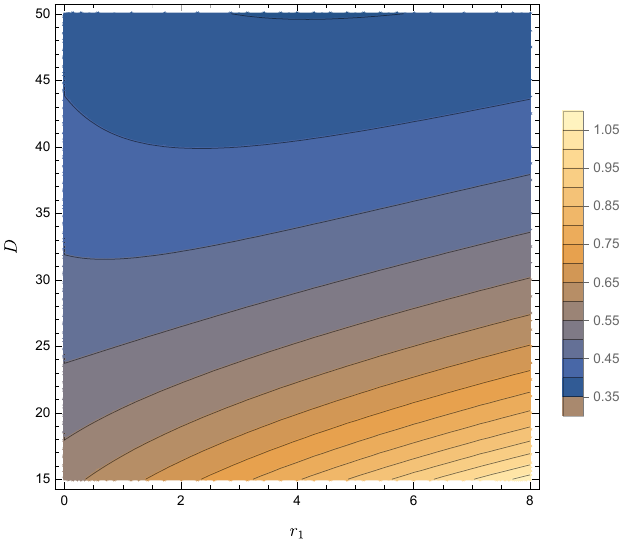}%
            \label{subfig:contour_d_smooth}%
        }
        \caption{(1d smooth potential / piecewise reset) Contour plots for the MFPT, $T_0$ in Eq.~\eqref{eq:smooth_sol_2r}, as a function of the reset rate $r_1$ and diffusion coefficient $D$, for the smooth quadratic potential $V(x) = x^2$ and the piecewise-constant reset rate defined in Eq.~\eqref{eq:reset_rates_pwquadratic}. Four different values $c \in \{ 0.01, 0.1, 1, 5 \}$ are considered in scaling the reset rate $r_2=cr_1$. Here we have fixed $L=0.01$, $x_0=4$ and $\beta=1$. All numerical evaluations were done using Mathematica \cite{Mathematica}.
        }
        \label{fig:contour_piecewiseresets}
    \end{figure*}

In Fig. \ref{fig:contour_piecewiseresets}, we further see that the benefit of resetting is dependent on the signal-to-noise ratio, in the sense that there is no optimal reset rate when the diffusion constant $D$ is small enough, $D\lesssim 25$.
We also observe that larger changes in MFPT behaviour occur when switching from a constant to a piecewise-constant reset rate with $r_2 > r_1$ than when switching to a piecewise-constant reset rate with $r_2 < r_1$.

\subsection{Non-smooth piecewise quadratic potential: constant and piecewise-constant reset rates}
\begin{figure}[tp]
    \centering
    \includegraphics[width=0.3\linewidth]{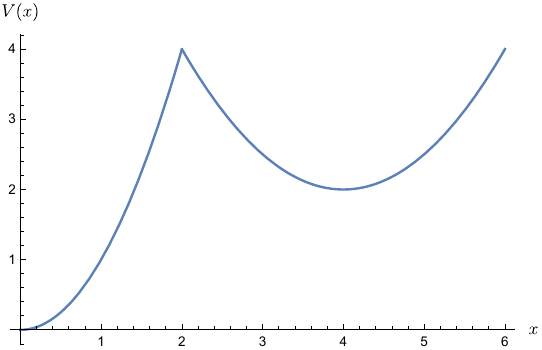}
    \caption{The piecewise quadratic potential $V(x)$ studied in this paper. See Eq. \eqref{eq:piecewisepot_1d} for the definition.}
    \label{fig:graphPiecewise}
\end{figure}

Let us now turn to the second case of interest to us, namely that of having a piecewise-polynomial potential. As a working example, we will consider the piecewise-quadratic potential
\begin{equation}\label{eq:piecewisepot_1d}
    V(x)=\begin{cases}
        x^2,\quad &x\leq 2,\\
        2+\frac12(x-4)^2,\quad &x>2.
    \end{cases}
\end{equation}
This has two minima, a global one at $x=0$ and a local one at $x=4$. See Fig.~\ref{fig:graphPiecewise}.

We now have the two equations
\begin{equation*}
    -1=\begin{cases}D\frac{\partial^2T(x)}{\partial x^2}-2x\frac{\partial T(x)}{\partial x}-r(x)T(x)+r(x)T_0,\quad &\text{for }x<2,\\
    D\frac{\partial^2T(x)}{\partial x^2}-(x-4)\frac{\partial T(x)}{\partial x}-r(x)T(x)+r(x)T_0,\quad &\text{for }x>2.\end{cases}
\end{equation*}

Let us again look at the two cases separately, i.e., having a constant reset rate or having two different reset rates. 

\medskip

\paragraph{Constant reset:} The general solutions are
\begin{equation}\label{eq:gen1Piecewise}
    T_1(x)=A_1H_{-\tfrac{r}{2}}\left(\frac{x}{\sqrt{D}}\right)+B_1\,{}_1F_1\left[\tfrac{r}{4},\tfrac12,\tfrac{x^2}{D}\right]+\frac{1}{r}+T_0,\quad x<2
\end{equation}
and
\begin{equation}\label{eq:gen2Piecewise}
    T_2(x)=A_2H_{-r}\left(\frac{x-4}{\sqrt{2D}}\right)+B_2\,{}_1F_1\left[\tfrac{r}{2},\tfrac12,\tfrac{(x-4)^2}{2D}\right]+\frac{1}{r}+T_0,\quad x>2,
\end{equation}
where $A_1, A_2, B_1,$ and $B_2$ are scalar constants to be determined.   
The boundary conditions are that $T(x)$ should be finite for $x\to\infty$, $T(L)=0$ and both $T(x)$ and $T'(x)$ should be continuous at the boundary of the two pieces $x=2$. This fixes all the constants $A_j$, $B_j$ and the final solution will again depend on whether we assume that $L$ and $x_0$ are larger than or smaller than 2. 

\begin{figure}[tp]
\centering
    \begin{subfigure}{.48\linewidth}
    \includegraphics[width=\linewidth]{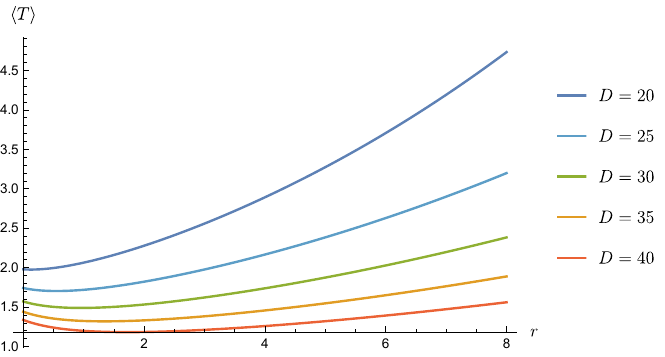}
    \caption{}
    \label{subfig:piecewise_constant_a}
    \end{subfigure}
    \hfill
    \begin{subfigure}{.48\linewidth}
    \includegraphics[width=.7\linewidth]{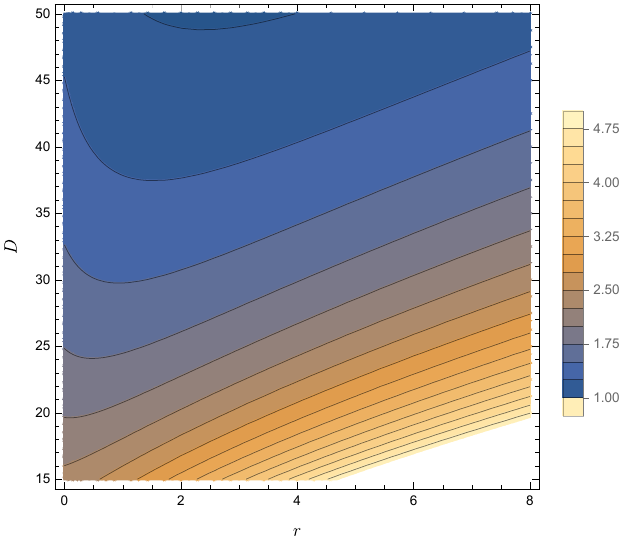}
    \caption{}
    \label{subfig:piecewise_constant_b}
    \end{subfigure}
\caption{ (1d non-smooth potential / constant reset / varying $D$) MFPT, $\langle{T}\rangle \coloneqq T_0$, of reaching a target $x=L$ for the non-smooth piecewise quadratic potential defined in Eq.~\eqref{eq:piecewisepot_1d}, computed using Eq.~\eqref{eq:nonsmooth_sol_constantR}.
Here we have fixed $L=0.01$ and $x_0=6$. \newline
(a) MFPT as a function of the constant reset rate $r$ for values of $D \in \{ 20, 25, 30, 35, 40 \}$.
(b) Contour plot of MFPT as a function of $r$ and $D$.
All numerical evaluations were done using Mathematica \cite{Mathematica}.
}
\end{figure}

As before, we will focus on the case where $L<2$ and $x_0>2$, i.e., where we start (and reset) in the valley of the local minimum, but the target is in the valley of the global one. After some algebraic manipulation, we find the solution of the MFPT for the non-smooth piecewise quadratic potential and constant reset rate:
\begin{equation}\label{eq:nonsmooth_sol_constantR}\footnotesize
    \begin{aligned}
        T(x_0)=&\left( rH_{-r}\left(\tfrac{x_0-4}{\sqrt{2D}}\right)\left(2H_{-\tfrac{r}{2}}\left(\tfrac{2}{\sqrt{D}}\right)\,{}_1F_1\left[1+\tfrac{r}{4},\tfrac{3}{2},\tfrac{4}{D}\right]+\sqrt{D}H_{-1-\tfrac{r}{2}}\left(\tfrac{2}{\sqrt{D}}\right)\,{}_1F_1\left[\tfrac{r}{4},\tfrac{1}{2},\tfrac{4}{D}\right]\right)\right)^{-1}\\
        \times&\Big(-H_{-r}\left(\tfrac{x_0-4}{\sqrt{2D}}\right)\left(2H_{-\tfrac{r}{2}}\left(\tfrac{2}{\sqrt{D}}\right)\,{}_1F_1\left[1+\tfrac{r}{4},\tfrac{3}{2},\tfrac{4}{D}\right]+\sqrt{D}H_{-1-\tfrac{r}{2}}\left(\tfrac{2}{\sqrt{D}}\right)\,{}_1F_1\left[\tfrac{r}{4},\tfrac{1}{2},\tfrac{4}{D}\right]\right)\\
        &+H_{-r}\left(-\sqrt{\tfrac{2}{D}}\right)\left(2H_{-\tfrac{r}{2}}\left(\tfrac{L}{\sqrt{D}}\right)\,{}_1F_1\left[1+\tfrac{r}{4},\tfrac32,\tfrac{4}{D}\right]+\sqrt{D}H_{-1-\tfrac{r}{2}}\left(\tfrac{2}{\sqrt{D}}\right)\,{}_1F_1\left[\tfrac{r}{4},\tfrac12,\tfrac{L^2}{D}\right]\right)\\
        &+\sqrt{2D}H_{-1-r}\left(-\sqrt{\tfrac{2}{D}}\right)\left(H_{-\tfrac{r}{2}}\left(\tfrac{L}{\sqrt{D}}\right)\,{}_1F_1\left[\tfrac{r}{4},\tfrac12,\tfrac{4}{D}\right]-H_{-\tfrac{r}{2}}\left(\tfrac{2}{\sqrt{D}}\right)\,{}_1F_1\left[\tfrac{r}{4},\tfrac{1}{2},\tfrac{L^2}{D}\right]\right)\Big).
    \end{aligned}
\end{equation}
This is plotted as a function of $r$ in Fig.~\ref{subfig:piecewise_constant_a}, for fixed $x_0=6$, $L=0.01$, and for the same values of $D$ considered in the smooth quadratic potential case. As before, we see that for a sufficiently large diffusion coefficient $D$, there is an optimal reset rate. It is noteworthy that the signal-to-noise ratio, defined by $|V'(x_0)|/\sqrt{4D}$ as before, is four times larger than in the smooth quadratic potential case, because $V'(x_0) = 2$ instead of $8$.
Despite this, the MFPT curves are similar to those in the smooth case, because the algorithm needs to overcome the local optimum valley centered at $x=4$ in order to reach the global optimum at $x=0$. It is clear that a higher noise level is needed to reach the global minimum in the non-smooth case, compared to the smooth case. In Fig.~\ref{subfig:piecewise_constant_b}, we also see that the benefits of resetting disappear when the diffusion coefficient is small enough, $D\lesssim 20$.

\medskip

\begin{figure}[tp]
\centering
\begin{subfigure}{.48\linewidth}
    \includegraphics[width=\linewidth]{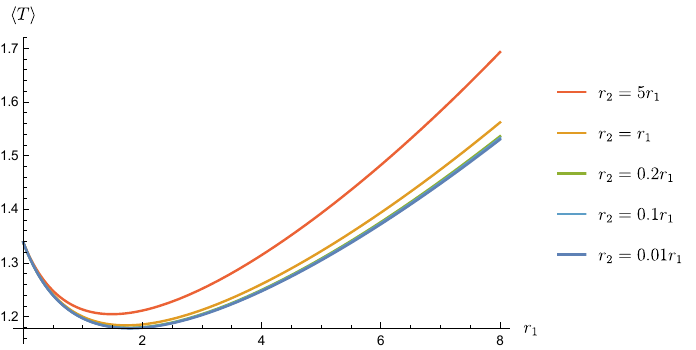}
    \caption{}
    \label{subfig:compare_nonsmooth1_a}
    \end{subfigure}
    \hfill
    \begin{subfigure}{.48\linewidth}
    \includegraphics[width=\linewidth]{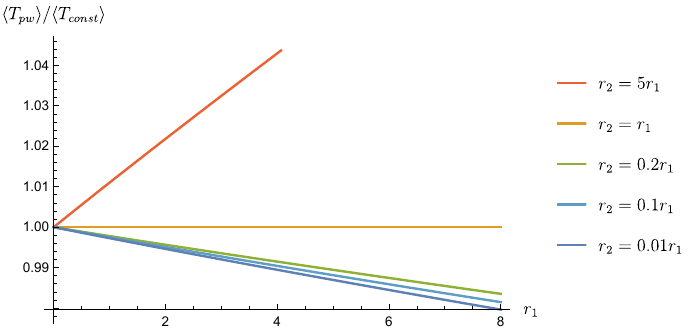}
    \caption{}
    \label{subfig:compare_nonsmooth1_b}
    \end{subfigure}
\caption{(1d non-smooth potential / piecewise reset / $D$ fixed) Comparison of the MFPT, $\langle{T}\rangle \coloneqq T_0$ in Eq.~\eqref{eq:nonsmooth_sol_2r}, between using a constant reset rate and the piecewise-constant reset rate defined in Eq.~\eqref{eq:2resets}, for the non-smooth piecewise-quadratic potential defined in Eq.~\eqref{eq:piecewisepot_1d}.
Here we have fixed $L=0.01$, $x_0=6$, $\beta=1$, and $r_2$ is set as a fixed multiple of $r_1$, i.e., the MFPT is plotted as a function of $r_1$ with $r_2=cr_1$, for values of $c \in \{ 0.01, 0.1, 0.2, 1, 5 \}$. \newline
(a) MFPT, labelled by $c$. 
(b) Ratio between the piecewise-constant reset rate, $\langle{T_{pw}}\rangle$, and the constant reset rate, $\langle{T_{const}}\rangle$, labelled by $c$. 
All numerical evaluations were done using Mathematica \cite{Mathematica}.
}
\label{fig:compare_nonsmooth1}
    \end{figure}

\begin{figure}[tp]
\centering
    \begin{subfigure}{.48\linewidth}
    \includegraphics[width=\linewidth]{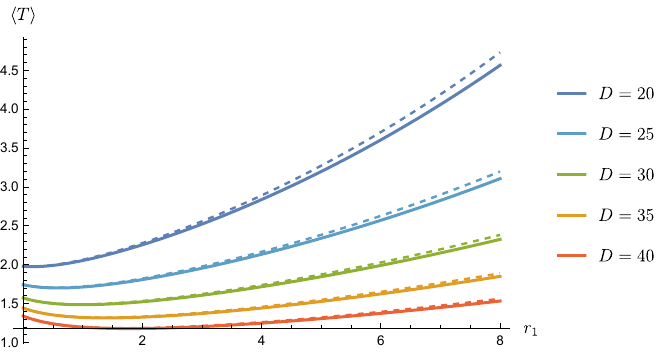}
    \caption{$r_2 = 0.1 r_1$}
    \label{subfig:compare_nonsmooth2_a}
    \end{subfigure}
    \hfill
    \begin{subfigure}{.48\linewidth}
    \includegraphics[width=\linewidth]{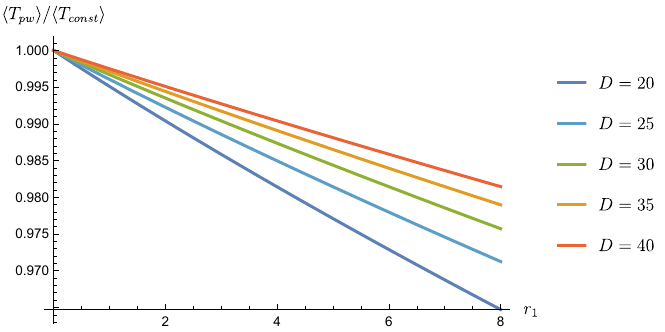}
    \caption{$r_2 = 0.1 r_1$}
    \label{subfig:compare_nonsmooth2_b}
    \end{subfigure}
\caption{(1d non-smooth potential / piecewise reset / varying $D$) 
Comparison of the MFPT, $\langle{T}\rangle \coloneqq T_0$ in Eq.~\eqref{eq:nonsmooth_sol_2r}, between using a constant reset rate and the piecewise-constant reset rate defined in Eq.~\eqref{eq:2resets} with $r_2 = 0.1 r_1$, for the non-smooth piecewise-quadratic potential defined in Eq.~\eqref{eq:piecewisepot_1d}, for values of $D \in \{20, 25, 30, 35, 40\}$. 
Here we have fixed $L=0.01$, $x_0=6$, $\beta=1$. \newline
(a) MFPT for constant reset rate ($\langle{T_{const}}\rangle$, dashed) and piecewise-constant reset rate ($\langle{T_{pw}}\rangle$, solid), labelled by $D$. (b) Ratio $\langle{T_{pw}}\rangle / \langle{T_{const}}\rangle$, labelled by $D$.
All numerical evaluations were done using Mathematica \cite{Mathematica}.
}
\label{fig:compare_nonsmooth2}
    \end{figure}

\begin{figure*}[tp]
        \subfloat[$r_2=5r_1$]{%
            \includegraphics[width=.48\linewidth]{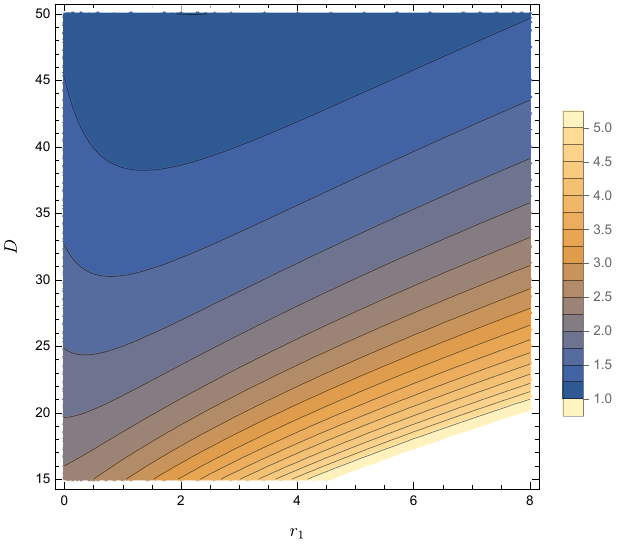}%
            \label{subfig:contour_a_nonsmooth}%
        }\hfill
        \subfloat[$r_2=r_1$]{%
            \includegraphics[width=.48\linewidth]{figures/T0_constant_contour.pdf}%
            \label{subfig:contour_b_nonsmooth}%
        }\\
        \subfloat[$r_2=0.1r_1$]{%
            \includegraphics[width=.48\linewidth]{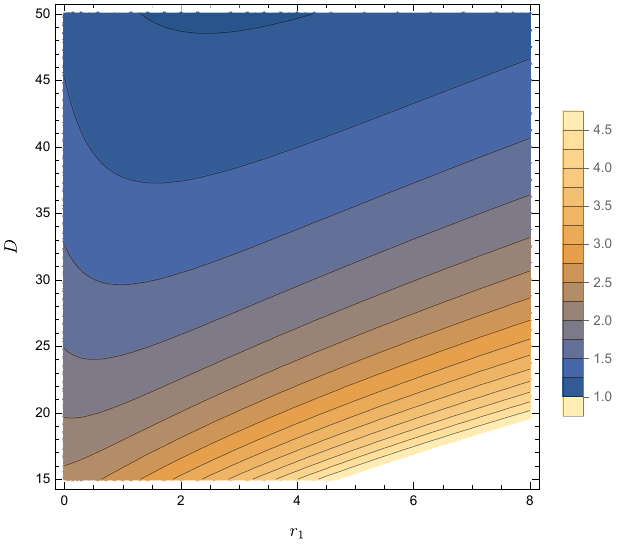}%
            \label{subfig:contour_c_nonsmooth}%
        }\hfill
        \subfloat[$r_2=0.01r_1$]{%
            \includegraphics[width=.48\linewidth]{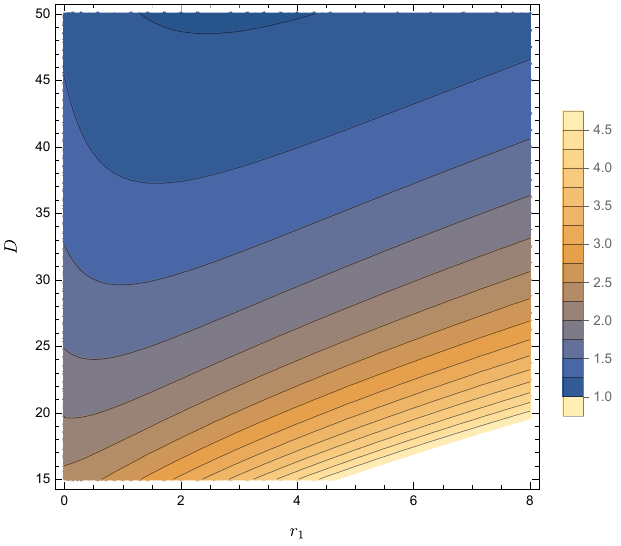}%
            \label{subfig:contour_d_nonsmooth}%
        }
        \caption{(1d non-smooth potential / piecewise reset) Contour plot for the MFPT, $T_0$ in Eq.~\eqref{eq:nonsmooth_sol_2r}, as a function of the reset rate $r_1$ and diffusion coefficient $D$, for the non-smooth piecewise quadratic potential defined in Eq.~\eqref{eq:piecewisepot_1d} and the piecewise-constant reset rate defined in Eq.~\eqref{eq:reset_rates_pwquadratic}.
        Four different $r_2=cr_1$ values are considered, for selected values of $c \in \{ 0.01, 0.1, 1, 5 \}$.
        Here we have fixed $L=0.01$, $x_0=6$ and $\beta=1$.
        All numerical evaluations were done using Mathematica \cite{Mathematica}.
        }
        \label{fig:contour_piecewiseresets_piecewisePot}
    \end{figure*}

\paragraph{Piecewise-constant reset rate}
We can again consider the case where we switch between two different reset rates $r_1$ and $r_2$
as in Eq. \eqref{eq:2resets}. There are now many options for the condition function $f(x)$. As before, one natural condition is that it depends on the gradient of the potential, i.e.,
\begin{equation*}
    f(x)=V'(x)=\begin{cases}
        2x,\quad &x\leq 2,\\
        x-4,\quad &x>2.
    \end{cases}
\end{equation*}
This would mean that
\begin{equation}\label{eq:reset_rates_pwquadratic}
    r(x)=\begin{cases}
        r_1,\quad &|x|>\tfrac{\beta}{2},\,x\leq 2,\\
        r_2,\quad &|x|\leq\tfrac{\beta}{2},\,x\leq 2,\\
        r_1,\quad &|x-4|>\beta,\,x>2,\\
        r_2,\quad &|x-4|\leq \beta,\,x>2,
    \end{cases}
\end{equation}
and we have four different solutions to the differential equations to consider, namely Eqs. \eqref{eq:gen1Piecewise} and \eqref{eq:gen2Piecewise}, each with the two different reset rates:
\begin{equation}\label{eq:Tpiecewise2rGeneral}
    T(x)=\begin{cases}
        A_1H_{-\tfrac{r_1}{2}}\left(\frac{x}{\sqrt{D}}\right)+B_1\,{}_1F_1\left[\tfrac{r_1}{4},\tfrac12,\tfrac{x^2}{D}\right]+\frac{1}{r_1}+T_0,\quad &x\leq2,\,|x|>\beta/2,\\
        A_2H_{-\tfrac{r_2}{2}}\left(\frac{x}{\sqrt{D}}\right)+B_2\,{}_1F_1\left[\tfrac{r_2}{4},\tfrac12,\tfrac{x^2}{D}\right]+\frac{1}{r_2}+T_0,\quad &x\leq2,\,|x|\leq\beta/2,\\
        A_3H_{-r_1}\left(\frac{x-4}{\sqrt{2D}}\right)+B_3\,{}_1F_1\left[\tfrac{r_1}{2},\tfrac12,\tfrac{(x-4)^2}{2D}\right]+\frac{1}{r_1}+T_0,\quad &x>2,\,|x-4|>\beta,\\
        A_4H_{-r_2}\left(\frac{x-4}{\sqrt{2D}}\right)+B_4\,{}_1F_1\left[\tfrac{r_2}{2},\tfrac12,\tfrac{(x-4)^2}{2D}\right]+\frac{1}{r_2}+T_0,\quad &x>2,\,|x-4|\leq\beta.
    \end{cases}
\end{equation}
The boundary conditions are the same as before, finiteness at infinity, zero at the target, and continuous over the various break points. The final solution is rather tedious, so we produce the full expression only in App.~\ref{app:MFPT}. It does depend on assumptions on the values of $L$, $\beta$ and $x_0$, again. The general behaviour is similar to before, and we can compare this solution for various ratios between the reset rates. The results are equivalent to what we found in the smooth case, and are shown in Figs.~\ref{fig:compare_nonsmooth1}-\ref{fig:contour_piecewiseresets_piecewisePot}. We fix $L=0.01$, $x_0=6$, $\beta=1$ and consider $r_2=cr_1$. In Fig.~\ref{fig:compare_nonsmooth1}, we further fix $D=40$ and plot the MFPT as a function of $r_1$ for different ratios $r_2/r_1=c$. As in the smooth case, we see that it is beneficial to use the piecewise-resetting whenever $r_2<r_1$. In Fig.~\ref{fig:compare_nonsmooth2}, we vary diffusion coefficient $D$ and fix $r_2=0.1r_1$. Here, we see that the benefits of using the piecewise-resetting are larger when diffusion coefficient $D$ is smaller. Finally, in Fig.~\ref{fig:contour_piecewiseresets_piecewisePot}, we plot the MFPT as a function of both the resetting rate $r_1$ and the diffusion coefficient $D$. We can see that, for all $r_2/r_1$ ratios considered, the optimal reset strategy disappears when the diffusion coefficient goes below a certain threshold, around $D\lesssim20$.
As in the 1d smooth potential, we observe that larger changes in MFPT behaviour occur when switching from a constant to a piecewise-constant reset rate with $r_2 > r_1$ than when switching to a piecewise-constant reset rate with $r_2 < r_1$.

We further note that, although the MFPT values are considerably lower for the smooth quadratic potential compared to the non-smooth piecewise quadratic potential, the \textit{relative} impact on the MFPT of using a piecewise-constant reset rate over a constant reset rate is strikingly similar across the two potentials, as becomes apparent by comparing Fig.~\ref{subfig:compare_smooth1_b} with Fig.~\ref{subfig:compare_nonsmooth1_b} and Fig.~\ref{subfig:compare_smooth2_b} with Fig.~\ref{subfig:compare_nonsmooth2_b}.

\medskip

\paragraph{Higher dimensions:} It is of interest to consider how these questions generalise to higher dimensions. We will not explore this at length here, however, it is worthwile to discuss some of the possibile generalisations. We leave it for future research to explore this further. 

When working in $\mathbb{R}^d$, the MFPT $T(x,L)$ of reaching a target $x=L\in\mathbb{R}^d$ will satisfy the following diffusion equation
\cite{evans2011diffusion,evans2011diffusionA,evans2014diffusion,ahmad2019first}
\begin{equation*}
        -1=D\nabla^2T(x,L)-\nabla V(x)\cdot \nabla T(x,L)-r(x)T(x,L)+r(x)T(x_0,L),
\end{equation*}
with boundary condition $T(L,L)=0$. 

We again restrict our analysis to quadratic potentials of the form generalizing Eq. \eqref{eq:piecewisepot_1d}:
\begin{equation}\label{quad_pot}
        V(x)=a+b(\norm{x}-c)^2,
\end{equation}
for some constants $a,b,c$, or rather, more generally, piecewise quadratic potentials where each piece is given by \eqref{quad_pot} for different values of the constants, such that it still is continuous, but not smooth. Note that this is highly symmetric and, in particular, depends only on the radial coordinate $R=\norm{x}$. The diffusion equation for the MFPT then simplifies to
\begin{equation}\label{diff_T_gen}
        -1=D(\partial^2_RT(R)+\tfrac{d-1}{R}\partial_RT(R))-2b(R-c)\partial_RT(R)-r(R)T(R)+rT_0,
\end{equation}
where we introduced the notation $T_0\coloneqq T(R_0,L)$ as before and further removed the dependence on $L$ in the notation for $T$, i.e., defined $T(R)\coloneqq T(R,L)$. In the following, we will also often drop the dependence on $R$ and simply write $T=T(R)$.

In general, Eq.\eqref{diff_T_gen} is, of course, very difficult to solve. When $r(x)=r$ we can define the variables $w(R)=T(R)-T_0-1/r$ and $y=\sqrt{\tfrac{b}{D}}R$ to write equation \eqref{diff_T_gen} on the form
\begin{equation*}
        yw''+(1+\alpha-\beta y-2y^2)w'+((\gamma-2-\alpha)y-\frac12(\delta+\beta(1+\alpha)))w=0,
\end{equation*}
with 
\begin{equation*}
    \begin{aligned}
        \alpha&\coloneqq d-2,\\
        \beta&\coloneqq -2c\sqrt{\frac{b}{D}},\\
        \gamma&\coloneqq d-\frac{r}{b},\\
        \delta&\coloneqq 2c\sqrt{\frac{b}{D}}(d-1).
    \end{aligned}
\end{equation*}
This is the biconfluent Heun equation \cite{batola1982generalisation}. Unfortunately, it seems that the solutions to this equation when $\alpha\in\mathbb{Z}$ have not been studied in much detail. For example, the general solutions used in much of the literature  \cite{batola1982generalisation}, do not work in this situation.

However, if we consider a smooth quadratic potential around the origin, $V(R)=kR^2$, $k>0$, the situation simplifies. For a constant resetting rate, we can solve this analytically for any dimension. The procedure is the same as before using the boundary conditions together with some known relations for hypergeometric functions, we end up with
\begin{equation}
    T(R_0)=\frac{1}{r}\left(\frac{U[\tfrac{r}{4k}, \tfrac{d}{2};\tfrac{kL^2}{D}]}{U[\tfrac{r}{4k},\tfrac{d}{2};\tfrac{kR_0^2}{D}]}-1\right),
\end{equation}
where $U[\cdot,\cdot;\cdot]$ is Tricomi's confluent hypergeometric function \cite{Tricomi1947}. 

\begin{figure}[tp]
\centering
    \begin{subfigure}{.48\linewidth}
    \includegraphics[width=\linewidth]{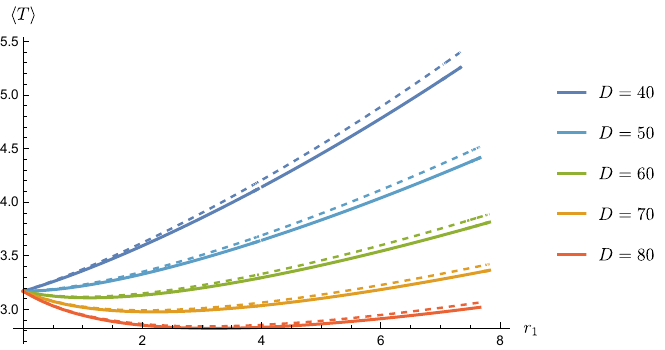}
    \caption{$r_2=0.1r_1$}
    \label{subfig:compare_nonsmooth22_a}
    \end{subfigure}
    \hfill
    \begin{subfigure}{.48\linewidth}
    \includegraphics[width=\linewidth]{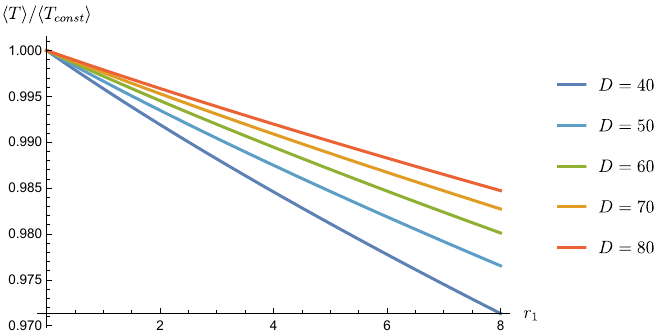}
    \caption{$r_2=0.1r_1$}
    \label{subfig:compare_nonsmooth22_b}
    \end{subfigure}
\caption{(2d smooth potential / piecewise reset / varying $D$) 
Comparison of the MFPT, $\langle{T}\rangle \eqqcolon T_0$ of Eq. \eqref{eq:MFPT2D}, between using a constant reset rate and the piecewise-constant reset rate defined in Eq.~\eqref{eq:resetCon_multidim} with $r_2 = 0.1 r_1$, for a 2-dimensional smooth quadratic potential $V(R)=R^2$, for values of $D \in \{40, 50, 60, 70, 80\}$.
Here we have fixed $L=0.01$, $R_0=4\sqrt{2}$, $\beta=1$. \newline
(a) MFPT for constant reset rate ($\langle{T_{const}}\rangle$, dashed) and piecewise-constant reset rate ($\langle{T_{pw}}\rangle$, solid), labelled by $D$. (b) Ratio $\langle{T_{pw}}\rangle / \langle{T_{const}}\rangle$, labelled by $D$.
All numerical evaluations were done using Mathematica \cite{Mathematica}.
}
\label{fig:compare_2D_r2_01r1}
    \end{figure}

When we consider the piecewise-constant reset rate, the solution becomes more complicated. We can simplify the analysis by focusing on a fixed dimension. Let us concentrate on $d=2$. We again introduce a condition for the switch between rates. The natural generalisation of our previous condition is
\begin{equation}\label{eq:resetCon_multidim}
    r(x)=\begin{cases}r_1,\quad  \norm{\nabla V(x)}> \beta,\\
    r_2,\quad \norm{\nabla V(x)}\leq \beta.\end{cases}
\end{equation}
The solution for the MFPT is again quite messy, so we produce the full expression only in App. \ref{app:MFPT2D}. There are some seemingly singular points, for example for $r_2=r_1=4$, this is however only an artefact of the particular solution techniques used and would not be a reflection of the actual MFPT. The results are similar as in the one-dimensional case, e.g., we again see that the benefit of using multiple reset rates is present when $r_2<r_1$. This is highlighted in Fig \ref{fig:compare_2D_r2_01r1}. We leave a more in depth analysis of the multi-dimensional case for future research.

\section{Numerical experiments}
\label{sec:results_experiments}

In order to corroborate the theoretical results obtained in the previous section, we consider a time-discretized version of the diffusion equation \eqref{eq:diffusion_equation} using the Euler-Maruyama approximation \cite{mou2019improvedboundsdiscretizationlangevin}, which is used to run the numerical experiments. The Euler-Maruyama approximation gives rise to a Markov process in $\mathbb{R}^d$, $\{x_k := x(k\Delta t)\}_{k\in\mathbb{N}}$ which, given the start value $x_0$, is defined by:
\begin{equation}\label{eq:diffusion_discrete}
    x_{k+1} = x_k + \mu F(x_k)\Delta t  + \sqrt{2D\Delta t} z_k,
\end{equation}
where $\Delta t$ is the step size resulting from a positive time discretization interval, and $\{z_k\}_{k \in \mathbb{N}}$ is an uncorrelated standard Gaussian process in $\mathbb{R}^d$, independent of the $\{x_k\}$ process.
As before, $F(x) = -\nabla V(x)$ where $V(x)$ is some potential in $\mathbb{R}^d$, and we set $\mu = 1$.

Each experiment consists of using Eq.~\eqref{eq:diffusion_discrete} to generate the iterates $x_k$, except when a resetting event occurs, in which case $x_k$ is reset to $x_0$. In addition, when the target $L$ is reached, the value $k\Delta t$ is recorded as a first passage time and the pair $(k, x_k)$ is reset to $(0, x_0)$. If $r(x)$ is the reset rate as a function of the position, the reset event is defined at each iteration $k$ as a Bernoulli variable with event probability $r(x_k) \Delta t$.

We consider different hyperparameter settings in order to analyze their impact on the MFPT. Each setting defines an experiment, which is run five times for up to one million iterations each.
The same five seeds are used across experiments so that any observed differences are only due to the different hyperparameter values, not to different seeds.
The MFPT of each experiment is estimated as the average of all recorded first passage times by the target, across all replications, giving the following expression for the estimate of the MFPT:
\begin{equation*}
    \widehat{\text{MFPT}} = \frac{\sum_{i=1}^{I} \sum_{n=1}^{N_i} \tau_{i,n}}{\sum_{i=1}^{I}N_i},
\end{equation*}
where $N_i$ is the number of first passage times after reset for replication $i$ of $I$, and $\tau_{i,n}$ is the $n$-th first passage time in the $i$-th replication, computed as the number of iterations to reach the target since reset, multiplied by $\Delta t$.
The event of reaching the target is triggered by the condition that the absolute difference between the observed potential at time step $k$, $V(x_k)$, and the potential value at the target, $V(L)$, is less than a  tolerance defined separately for each case analyzed below.

As potential $V(x)$, we consider the functions analyzed in the theoretical analysis of Section~\ref{sec:results_theory}: the smooth quadratic potential in one and two dimensions, and the 1d non-smooth piecewise quadratic potential with its 2d analogue found by revolving the 1d potential around the third axis, to obtain a symmetric piecewise parabolic function.
In all cases, the step size is set to $\Delta t = 10^{-4}$ and the target is set to the position of the global optimum of the potential, i.e., $L = 0$. The reset rate function $r(x)$ is defined by the piecewise-constant function \eqref{eq:resetCon_multidim} for all cases, parameter $\beta = 1$, and parameters $r_1 \not = r_2$. Note that the constant reset rate strategy uses $r_1 = r_2$.
For the purpose of comparing results with those of  Section~\ref{sec:results_theory}, the effect on the MFPT with four different reset rates, $r_1 = 0.1, 1, 5, 10$, three reset rate ratios $r_2/r_1 = 0.1, 1, 10$, and five diffusion coefficients $D$ (to be defined in each case) is analyzed. The value of $x_0$ depends on the potential and is defined in each subsection below.

\subsection{1d experiments}
With the purpose of verifying the theoretical results of Figs.~\ref{fig:compare_smooth2} and \ref{fig:compare_nonsmooth2} in Section~\ref{sec:results_theory}, as well as Figs.~\ref{fig:compare_smooth3} and \ref{fig:compare_nonsmooth3}, the same five diffusion coefficients $D$ and reset position $x_0$ used therein were considered in the 1d experiments. This means $D = [20, 25, 30, 35, 40], x_0 = 4$ for the smooth potential and $x_0 = 6$ for the non-smooth potential. The only difference with the analytic figures is that the target $L$ is set to $0$ instead of $0.01$.

Only two parameters belong exclusively to the numerical experiments context: the step size $\Delta t$ (already fixed at $10^{-4}$), and the tolerance $\Delta V_\mathrm{tol}$, which is set to $0.01$ for the 1d potentials.

With this parameter setup, the discrete-time version of the signal-to-noise ratio, defined in Section~\ref{sec:problem} as $|V'(x_0)|/\sqrt{4D}$, becomes $\frac{|V'(x_0)| \sqrt{\Delta t}}{\sqrt{4D}}$ and varies from a lowest value of $\approx 0.006$ at $D = 40$ to a highest value of $\approx 0.009$ at $D = 20$.
We note that the discrete-time signal-to-noise ratio decreases as $\Delta t$ decreases, for a fixed value of $D$.

\subsection{2d experiments}
In order to compare the 2d experiments with the theoretical results of Sec.~\ref{sec:results_theory}, we consider the equivalent reset points of $x_0 = (4, 4)$ for the smooth quadratic potential case. For the non-smooth piecewise quadratic case we pick the point $x_0 = (6, 6)$. We further enforce the same initial signal-to-noise ratio as in the respective 1d counterparts. Out of the two parameters exclusive to the numerical experiments, $\Delta t$ and $\Delta V_\mathrm{tol}$, $\Delta t$ is kept equal to $10^{-4}$ as in the 1d case, but $\Delta V_\mathrm{tol}$ is increased 10-fold to $0.1$, considering the 2d excursions of the $x_k$ iterates have more wandering space available, which increases the difficulty of finding the target compared to the 1d case.

The condition on using the same signal-to-noise ratio as in the 1d experiments is enforced through the selected values of $D$ in these 2d experiments. This is done with the help of Tab.~\ref{tab:snr}, which shows the expressions for the signal-to-noise ratio (SNR) for all four potentials considered, computed at their respective reset point. The $D$ values to use for the 2d problems, given the $D$ values of the respective 1d problems, is obtained by equalling the two columns in each of the two rows of the table.

Fig.~\ref{fig:trajectory_2d} in App.~\ref{app:trajectory_2d} shows an example of a diffusion trajectory on the 2d non-smooth piecewise quadratic potential.

\begin{table}[ht!]
    \centering
    \begin{tabular}{c|c|c}
                SNR         & 1d & 2d   \\
        \hline
         \begin{tabular}{c} smooth quadratic \\
         (defined for 1d case in Section~\ref{sec:problem} with $k = 1$)\end{tabular} & $x_0\sqrt{\frac{\Delta t}{D}}$, $x_0 = 4$ & $||{x_0}||\sqrt{\frac{\Delta t}{D}}$, $x_0 = (4, 4)$ \\
         \hline
         \begin{tabular}{c} piecewise quadratic \\
         (defined for 1d case in Eq.~\eqref{eq:piecewisepot_1d}) \end{tabular} & $(x_0 - 4)\sqrt{\frac{\Delta t}{4D}}$, $x_0 = 6$ & $(||x_0|| - 4) \sqrt{\frac{\Delta t}{4D}}$, $x_0 = (6, 6)$
    \end{tabular}
    \caption{Signal-to-noise ratio (SNR := $\frac{\norm{\nabla V(x_0)}}{\sqrt{4D}}$) for the four different potentials $V(x)$ considered in the discrete-time diffusion process defined in Eq.~\eqref{eq:diffusion_discrete}, with $F(x) := -\nabla V(x)$ and $x_0$ the reset point.
    }
    \label{tab:snr}
\end{table}

\begin{figure}[tp]
    \subfloat[1d smooth quadratic, $r_2=10r_1$]{%
        \includegraphics[width=.5\linewidth]{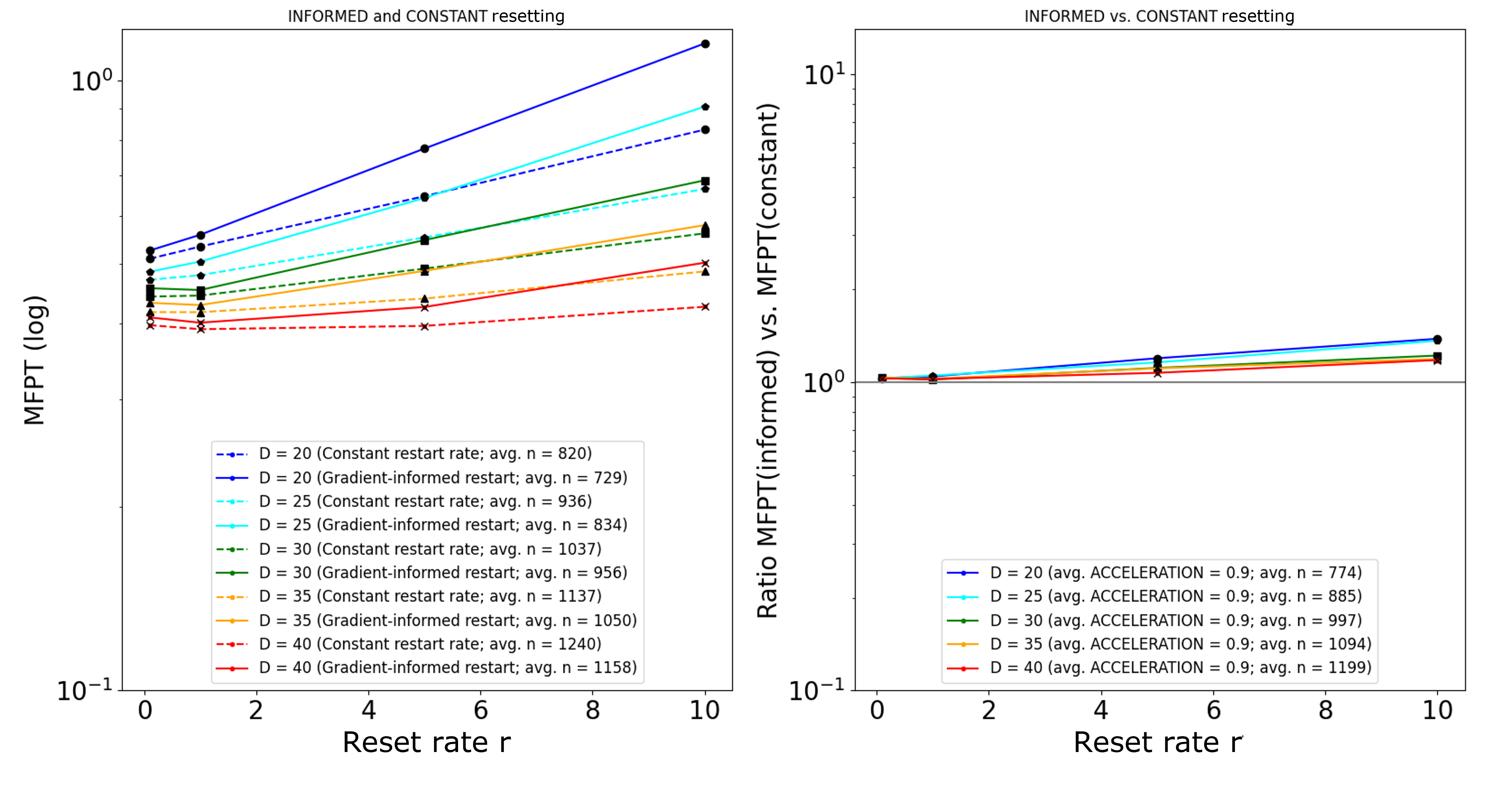}%
        \label{subfig:results_1d_a}%
    }\hfill
    \subfloat[1d smooth quadratic, $r_2=0.1r_1$]{%
        \includegraphics[width=.5\linewidth]{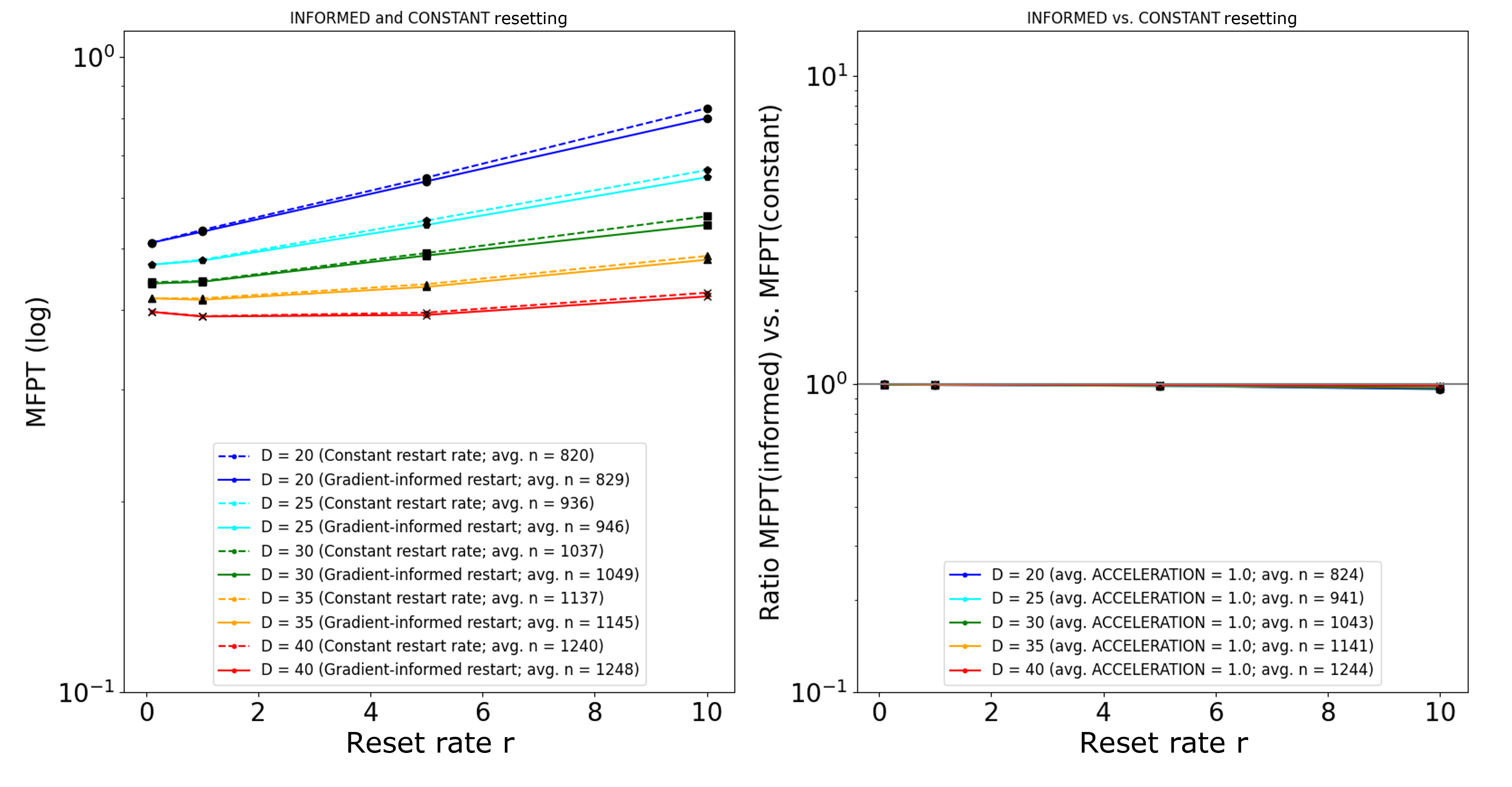}%
        \label{subfig:results_1d_b}%
    }\\
    \subfloat[1d piecewise quadratic, $r_2=10r_1$]{%
        \includegraphics[width=.5\linewidth]{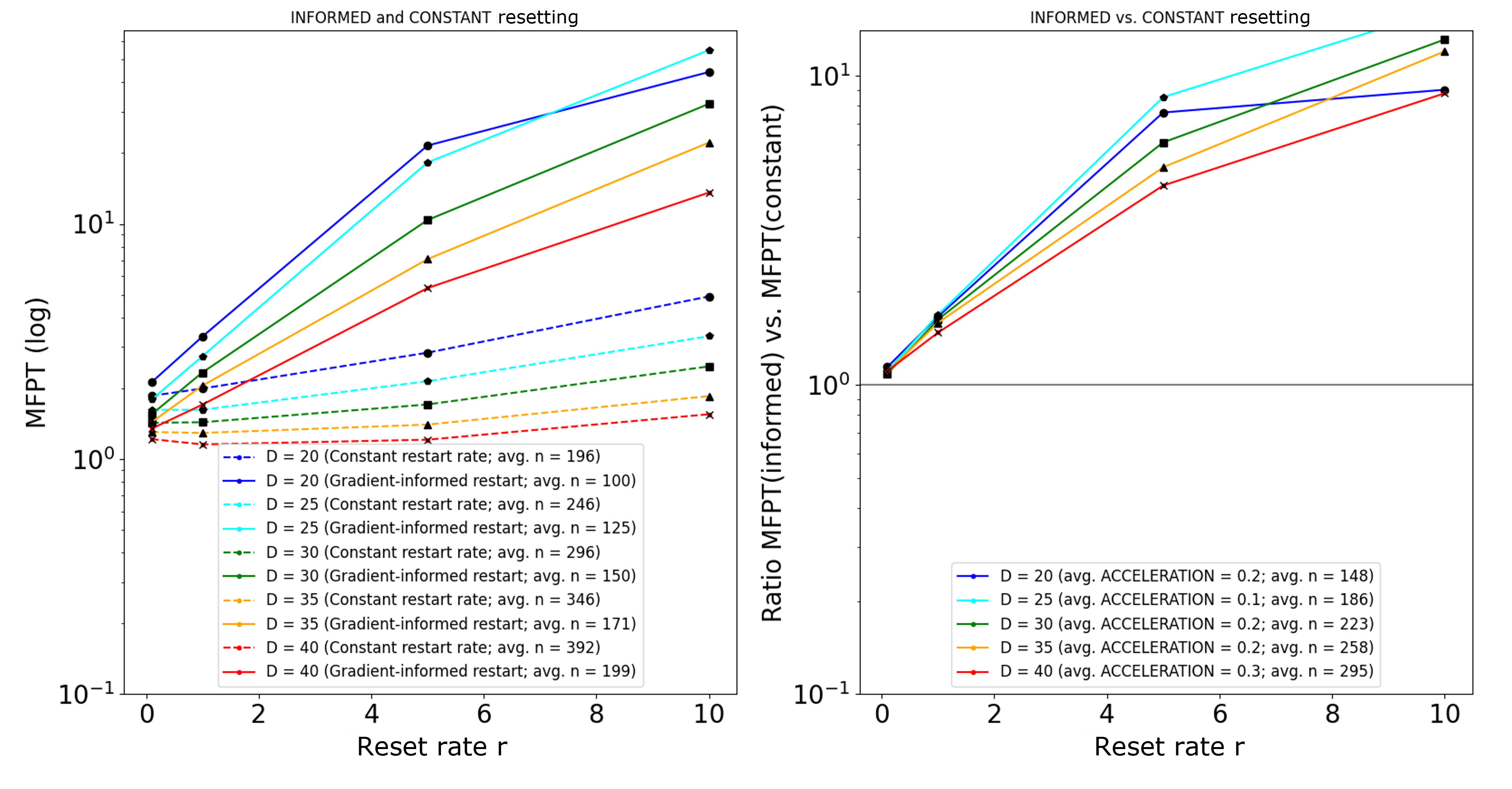}%
        \label{subfig:results_1d_c}%
    }\hfill
    \subfloat[1d piecewise quadratic, $r_2=0.1r_1$]{%
        \includegraphics[width=.5\linewidth]{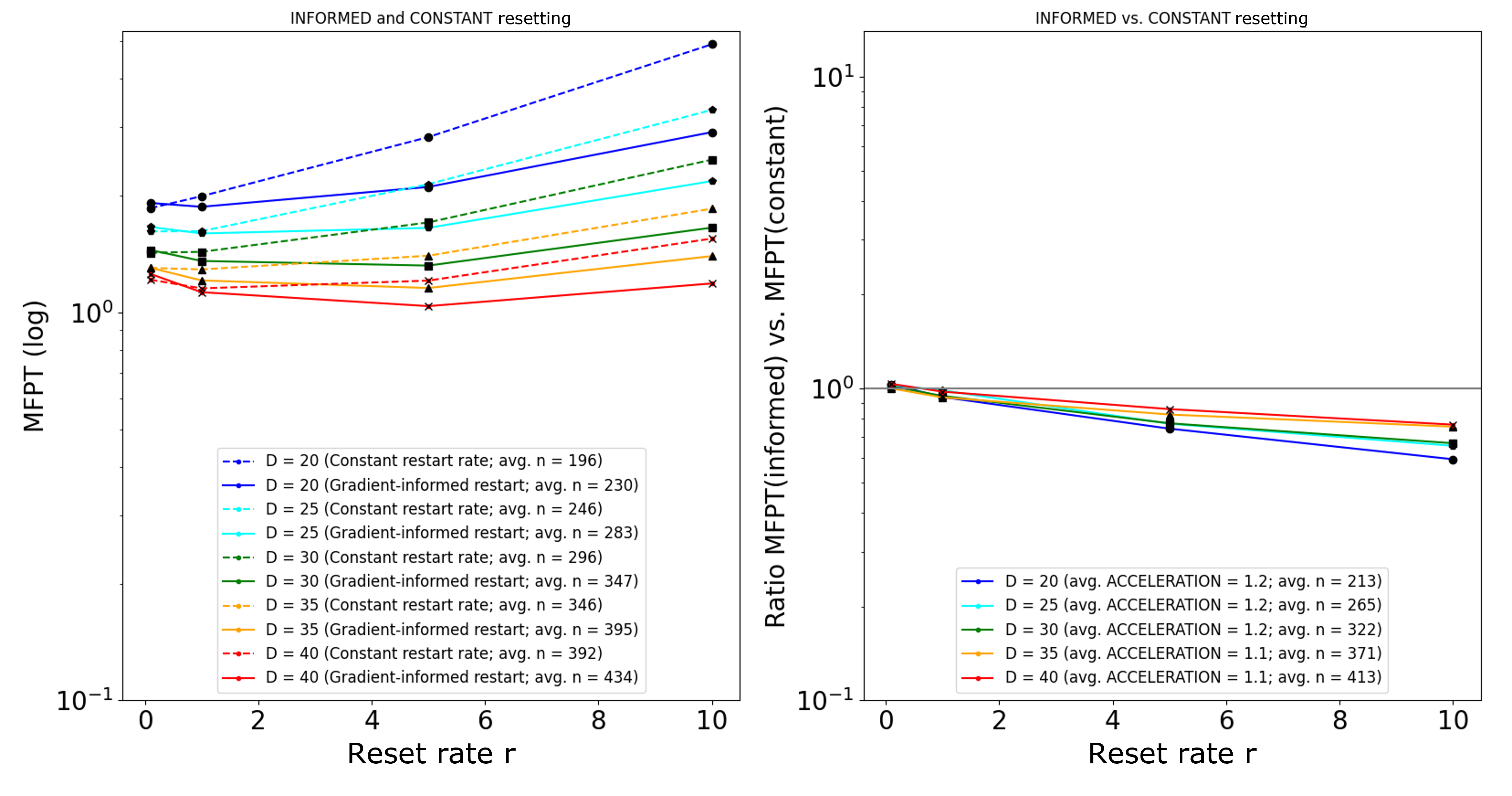}%
        \label{subfig:results_1d_d}%
    }
    \caption{Average MFPT (measured in time units) obtained on two 1d quadratic potentials, $V(x)$, using both the constant reset rate strategy and the piecewise-constant reset rate defined in Eq.~\eqref{eq:2resets} by setting $f(x) = V'(x)$. \newline
    (a) and (b): 1d smooth quadratic potential, $V(x) = x^2$. \newline
    (c) and (d): 1d non-smooth piecewise quadratic potential, with $V(x)$ defined in 
    Eq.~\eqref{eq:piecewisepot_1d}. \newline
    Reset rate $r_2$ is used when the gradient norm is smaller than the threshold $\beta = 1$, and otherwise reset rate $r_1$ is used.
    For visual comparison, all right plots across panels, showing the ratio between using a piecewise-constant reset rate vs. a constant reset rate, have the same vertical scales.
    }
\label{fig:results_1d}
\end{figure}

\begin{figure}[tp]
    \subfloat[2d smooth quadratic, $r_2=10r_1$]{%
        \includegraphics[width=.5\linewidth]{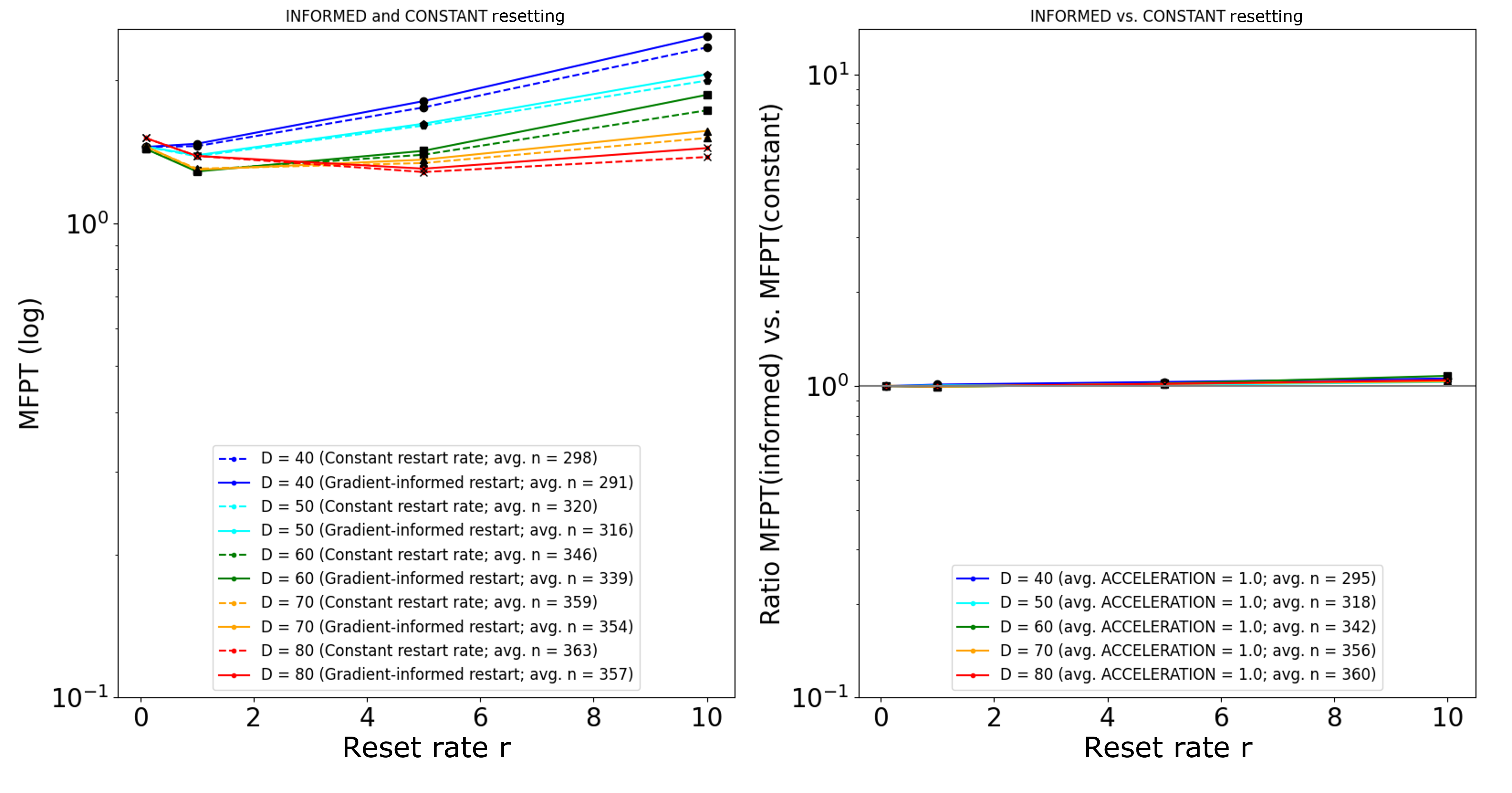}%
        \label{subfig:results_2d_a}%
    }\hfill
    \subfloat[2d smooth quadratic, $r_2=0.1r_1$]{%
        \includegraphics[width=.5\linewidth]{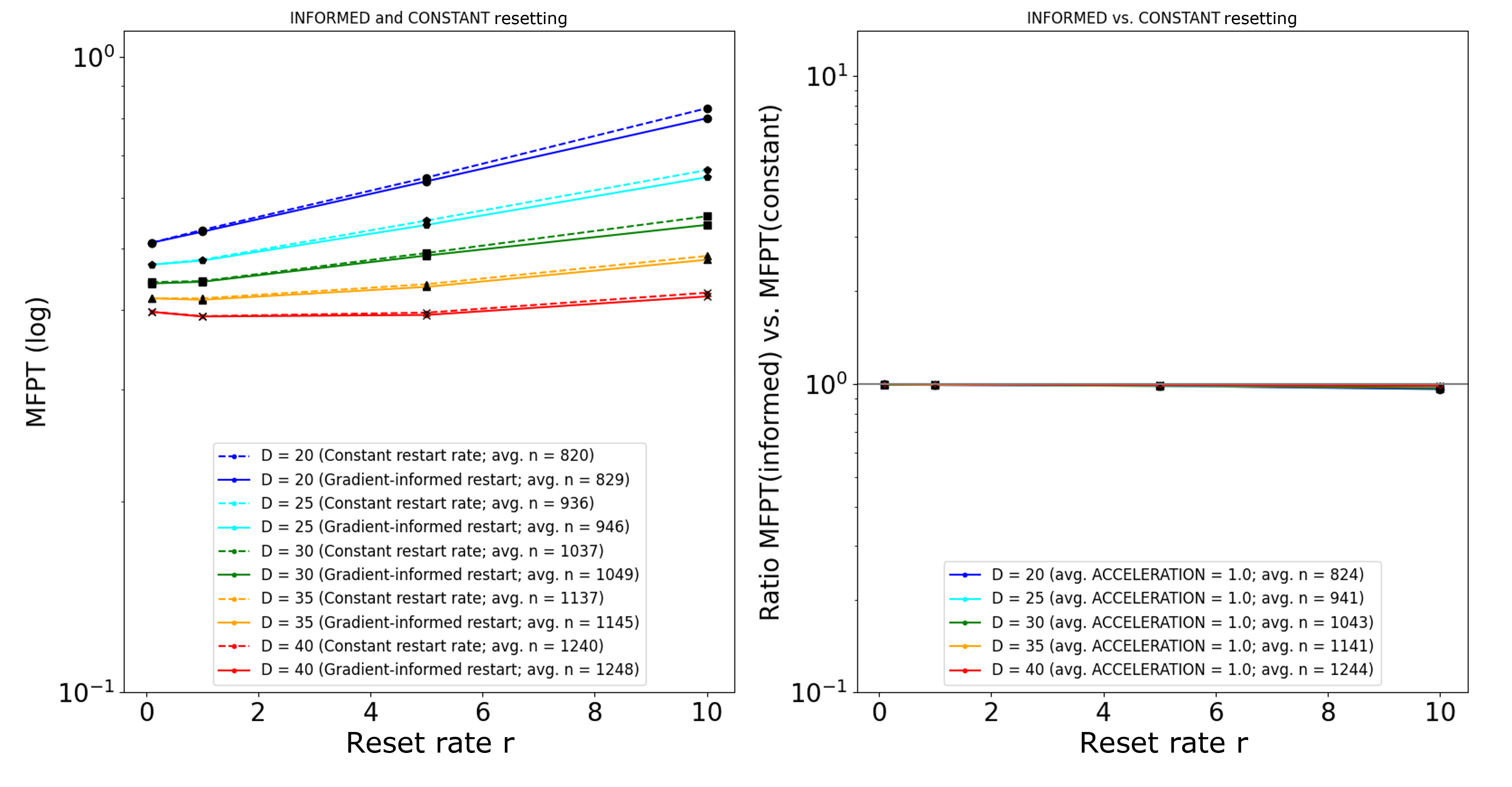}%
        \label{subfig:results_2d_b}%
    }\\
    \subfloat[2d piecewise quadratic, $r_2=10r_1$]{%
        \includegraphics[width=.5\linewidth]{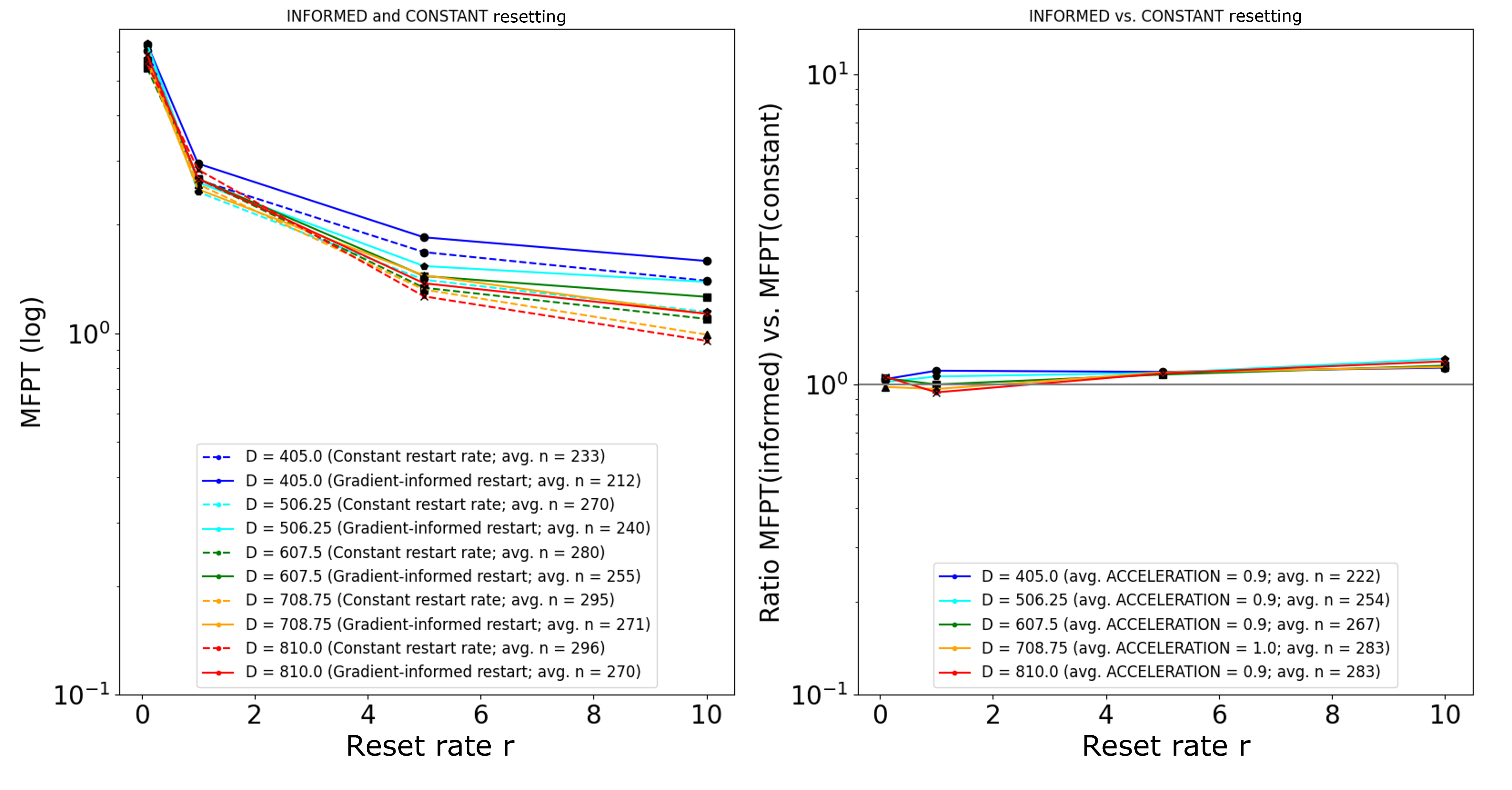}%
        \label{subfig:results_2d_c}%
    }\hfill
    \subfloat[2d piecewise quadratic, $r_2=0.1r_1$]{%
        \includegraphics[width=.5\linewidth]{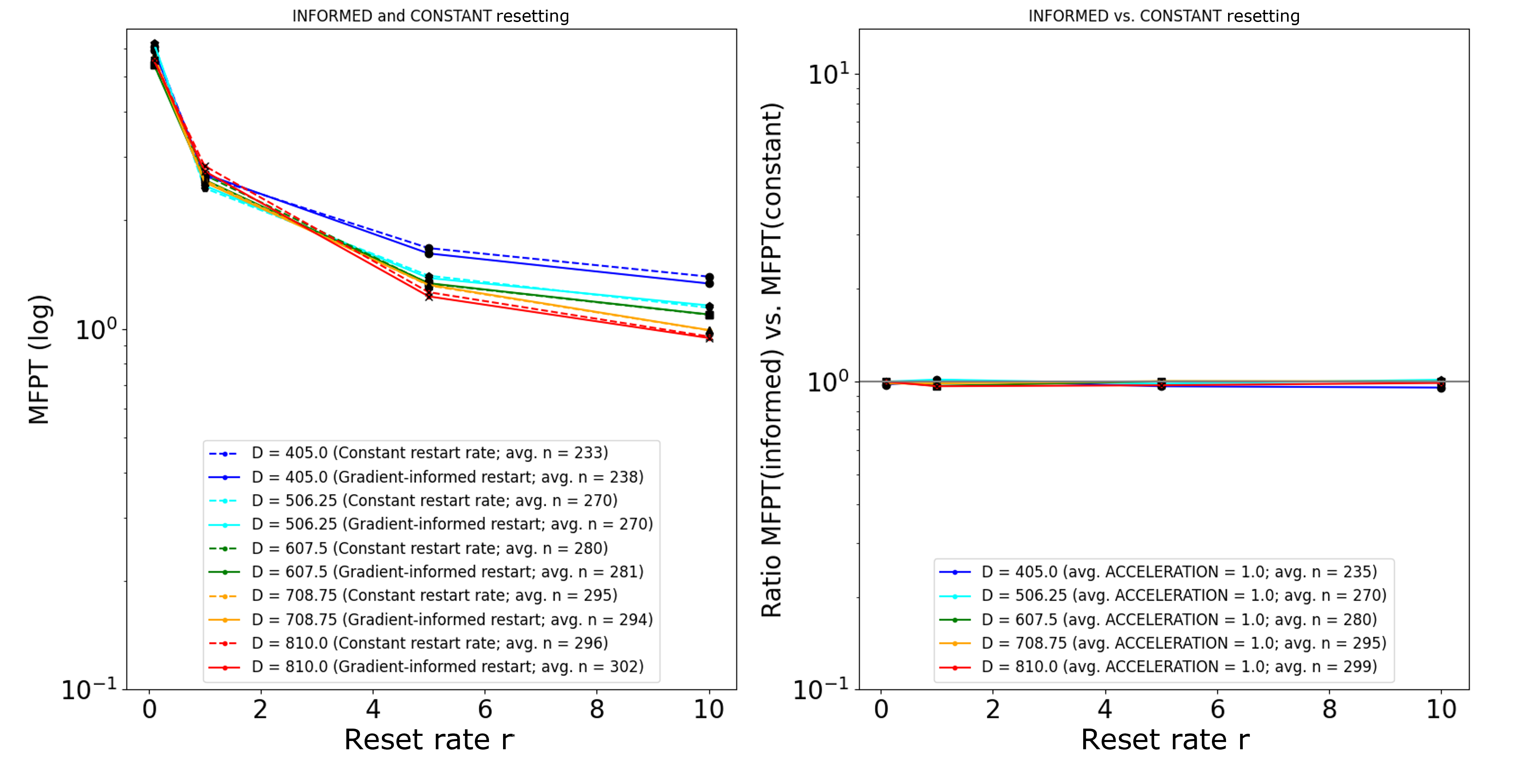}%
        \label{subfig:results_2d_d}%
    }
    \caption{Average MFPT (measured in time units) obtained on two 2d quadratic potentials, $V(x)$ with $x := (x_1, x_2)$, using either the constant reset rate or the piecewise-constant reset rate defined in \eqref{eq:resetCon_multidim}.
    Reset rate $r_2$ is used when the gradient norm is smaller than the threshold $\beta = 1$, and otherwise reset rate $r_1$ is used. \newline
    (a) and (b) smooth quadratic potential, $V(x_1, x_2) = x_1^2 + x_2^2$. \newline
    (c) and (d) non-smooth piecewise quadratic potential, obtained by revolving the $V(x)$ function defined in 
    Eq.\eqref{eq:piecewisepot_1d} around the $V$ axis. \newline
    For visual comparison, all right plots across panels, showing the ratio between using a piecewise-constant reset rate vs. a constant reset rate, have the same vertical scales.
    }
\label{fig:results_2d}
\end{figure}

\subsection{Analysis of results}
\label{sec:results_analysis}
Figs.~\ref{fig:results_1d} and \ref{fig:results_2d} present the results in terms of the MFPT as a function of the reset rate $r_1$ with curves labeled by $D$, respectively for the 1d and the 2d experiments. Each set of results is arranged in a $2 \times 2$ panel where the top and bottom rows correspond, respectively, to the smooth and non-smooth potentials, and the left and right columns correspond, respectively, to the reset rate ratios $r_2/r_1 = 10$ and $0.1$. The MFPT values obtained for the constant reset rate strategy are plotted as dashed curves in each of the four cells.

Following are the main observations that we can infer from the MFPT plots obtained for the numerical experiments:
\begin{enumerate}
    \item 1d numerical experiments confirm the following aspects of the theoretical results:
    \begin{enumerate}
        \item The MFPT tends to decrease as the diffusion coefficient $D$ increases, for a given reset rate.
        \item An optimum reset rate exists as long as the diffusion coefficient $D$ is large enough, i.e., for lower signal-to-noise ratios computed at the reset position.
        \item The curvature of the curve increases as the diffusion coefficient $D$ increases from 20 to 40, i.e., from blue to red curves.
        \item The regime $r_2 < r_1$ is better than $r_2 > r_1$ in terms of obtaining a smaller MFPT than in the constant reset rate strategy $r_2 = r_1$, for a given diffusion coefficient. This is verified, for $D = 40$, in Figs.~\ref{subfig:results_1d_a} and \ref{subfig:results_1d_b}, which confirm the theoretical behaviour plotted in Fig.~\ref{fig:compare_smooth1} for the smooth quadratic potential.  Figs.~\ref{subfig:results_1d_c} and \ref{subfig:results_1d_d} confirm the theoretical behaviour plotted in Fig.~\ref{fig:compare_nonsmooth1} for the non-smooth piecewise quadratic potential. 
        \item The relative impact on the MFPT of the piecewise-constant reset rate becomes larger as the diffusion coefficient $D$ decreases, for a given reset rate. This is true regardless of whether $r_2$ is smaller or larger than $r_1$, and is seen in Figs.~\ref{fig:results_1d}(a)-(d) for the numerical experiments, and respectively in Figs.~\ref{subfig:compare_smooth3_b}, \ref{subfig:compare_smooth2_b}, 
        \ref{subfig:compare_nonsmooth3_b}, \ref{subfig:compare_nonsmooth2_b} for the theoretical results.
    \end{enumerate}
    \item \label{itm:results_differences} However, differences are observed with the theoretical results in terms of the \textit{relative} decrease or increase in MFPT of using a piecewise-constant reset rate over a constant reset rate. The theoretical results presented in Figs.~\ref{subfig:compare_smooth1_b} and \ref{subfig:compare_nonsmooth1_b} show that the relative MFPT change is about \textit{the same} in both the smooth and non-smooth potentials. In contrast, the corresponding numerical results presented in Fig.~\ref{fig:results_1d} show that the relative MFPT change is \textit{larger} in the non-smooth 
    \medskip
    piecewise quadratic potential than in the smooth quadratic potential. \\
    We believe the difference is a consequence of both the time discretization of the original continuous-time problem and the tolerance hyperparameter $\Delta V_\mathrm{tol}$ used to signal the occurrence of reaching the target, as described above. More precisely, due to time discretization, there is no longer a continuous evolution of the $x(t)$ solution, but instead a sequence of random discrete steps $x_{k+1} - x_k$ computed according to Eq.~\eqref{eq:diffusion_discrete}. Therefore, the time to reach the target is clearly a function of the $\Delta V_\mathrm{tol}$ tolerance. Additionally, the particle may have actually \textit{passed by} the target from one particular $x_{k-1} \rightarrow x_k$ iteration, but with no detection of the target just because $V(x_k)$ does not satisfy the specified tolerance. Thus, the discrete-time algorithm needs to spend some time near the target before triggering a reset event, as this will increase the chances that $V(x_l)$ satisfies the tolerance at some future iteration $l$. This is why using a gradient-based piecewise-constant reset rate with $r_2 < r_1$ is beneficial, as observed both theoretically and numerically.
    However, in the discrete-time context,  this benefit is expected to be larger in the non-smooth piecewise-quadratic potential than in the smooth quadratic potential because a reset event sends the particle back to its original position, where the signal-to-noise ratio is smaller than in the smooth case (as mentioned previously), and the particle needs to overcome the local optimum valley before being able to reach the global optimum target.
    \item The 2d experiments show similar results as the 1d counterparts, but the benefits or hindrances of using a piecewise-constant reset rate are not as apparent as in the 1d case, even when the same initial signal-to-noise ratios are used in the respective cases.
\end{enumerate}

Finally, it is interesting to note that using a piecewise-constant reset rate with $r_2 > r_1$ can make the optimal reset rate (of the constant reset rate strategy) vanish. This can be clearly seen for the non-smooth piecewise quadratic potential in the left subplot of Fig.~\ref{subfig:results_1d_c}.

\section{Discussion}
In this paper, we have considered diffusions with drift given by the gradient of a potential function and with stochastic resetting to a fixed position. We have analyzed the mean first passage time (MFPT) for the scenario when the potential of the drift is non-smooth, in particular piecewise quadratic, as well as the scenario where the reset rate is piecewise-constant.

The theoretical analysis of the non-smooth case shows that the benefit of resetting is dependent on the interplay between drift and noise, e.g., when the drift at the reset position is weak (small potential) compared to the noise (large diffusion coefficient), the resetting is more likely to be beneficial. This is also consistent with previous literature on the smooth case \cite{ahmad2019first,ray2019peclet}. We further find that having a piecewise-constant reset rate is sometimes beneficial, but not always. This depends on the condition for switching between the different rates as well as the ratio between the rates. In our analysis, with a gradient dependent resetting rate, we found that it is detrimental to use a higher reset rate for smaller gradient regions, but beneficial to switch to a lower rate in those same regions (some of which may occur near the given target). See plots of theoretical curves in Figs.~\ref{fig:compare_smooth2} and \ref{fig:compare_nonsmooth1}, as well as the numerical experiment curves in Figs.~\ref{fig:results_1d} and \ref{fig:results_2d}. 
This may suggest iterative gradient-based optimisation algorithms with reset rates conditional on both the objective-function value and some norm of the subgradient at each iteration.

Additionally, we have confirmed with numerical experiments all theoretical results developed on the 1d case, except for the differences described in item~(\ref{itm:results_differences}) of Section~\ref{sec:results_analysis}, which regard the \textit{relative} impact on the MFPT of using a piecewise-constant reset rate compared to using a constant one. In the 1d numerical experiments, this relative change is stronger in the non-smooth piecewise quadratic potential than in the smooth quadratic potential, whereas it is the same from the theoretical standpoint. The conjectured reasons for this discrepancy are explained in Section~\ref{sec:results_analysis}. Although this discrepancy is less apparent in the 2d case, it suggests that space-dependent resetting may be more beneficial in practice than predicted by theory, especially in non-convex problems.

Clearly, the analysis and experiments of our paper are limited to very simple non-smooth potentials. In particular, the gradient threshold $\beta$ used to define the piecewise-constant reset rate has been set to $1$ throughout, based on the knowledge of the potentials analyzed.
In practical settings, where the potential function is generally not known but instead learned from data collected as the discrete diffusion algorithm progresses, sensible strategies to define the threshold are called for. For instance, the authors in \cite{mastropietro2026ieeequantum} consider a burn-in exploration period, with no reset events, which is used to estimate an average gradient value of which a fraction is used as the $\beta$ parameter defining the piecewise-constant resetting. Robustness against noise could be incorporated by choosing an appropriate quantile of the gradient norm distribution, albeit at higher storage costs. It could also be updated with a certain frequency based on updated gradient norm distributions, computed on a moving window. Information about the variability of the observed gradient values could also be leveraged, as done in many gradient-based algorithms, such as Nesterov's accelerated gradient method \cite{Nesterov1983NAG}.

It should be noted that a few optimisation algorithms typically used in machine learning, e.g. the Adam algorithm \cite{kingma2015adamiclr}, implement adaptive step sizes $\Delta t$ that are adjusted based on the past values of the function and its gradient. Even though such approaches could be framed into our context by keeping a constant $\Delta t$ value and absorbing any adjustments into a time-dependent mobility parameter $\mu$, they fall outside the scope of this work for two main reasons: (i) the $\mu$ parameter is assumed to be constant (equal to $1$) in the Chapman-Kolmogorov Eq.~\eqref{eq:chapman_kolmogorov}, and (ii) the system's condition after reset is not the same as at the onset of the optimisation algorithm because new function and gradient information has been collected by then and thus impact the mobility parameter $\mu$ after reset. The theoretical analysis would therefor need adjusting in order to account for such settings. This would be an interesting direction for future research.

In sum, as seen by the results of this paper which confirm the classic results of Evans and Majumdar, \cite{evans2011diffusion}, resetting can either benefit or hinder the search depending on the parameters of the process. Pal et al., \cite{pal2019landau}, developed a Landau-like theory to characterise these phase transitions in resetting processes. This is further studied in \cite{ahmad2022first}. It would be interesting to study these aspects further for the settings discussed in our paper. It would also be of interest to explore \textit{restarting} strategies, i.e., where the reset position is selected at random in the solution space, as this will promote exploration of other regions and thus may help avoid local optima and find the global optimum faster. We plan to explore this approach in future work as well as to generalize the choice of the gradient threshold for practical optimisation problems where the potential function is unknown.

\vspace*{-0.5cm}

\begin{section}{Acknowledgements} 
	We are grateful to the two anonymous reviewers, whose comments helped improve the paper. This work has received funding from the European Union’s Horizon Europe research and innovation programme under grant agreement No. 101070568.
\end{section}

\appendix

\section{The MFPT for piecewise quadratic potential with two reset rates}\label{app:MFPT}
For completeness, we give the full expression for the solutions of the MFPT in the case when we have a one-dimensional piecewise quadratic potential and two different reset rates. In other words, the solution to the boundary value problem in \eqref{eq:Tpiecewise2rGeneral}. The exact solution depends on the assumptions on the values of the target $L$, the cutoff $\beta$ and the starting position $x_0$. The case of most interest, which is also the one studied in the main text is for $0<\beta\leq1$, $L\leq \tfrac{\beta}{2}$ and $x_0>4$. With these conditions we find the expression of the MFPT for the non-smooth piecewise quadratic potential with piecewise-constant reset rate from the boundary value problem \eqref{eq:Tpiecewise2rGeneral}:

\begin{equation}\label{eq:nonsmooth_sol_2r}
\footnotesize
    \begin{aligned}
        T_0=&\frac{1}{r_1r_2H_{-r_1}\left(\tfrac{x_0-4}{\sqrt{2D}}\right)\left(2H_{-\tfrac{r_1}{2}}(\tfrac{2}{\sqrt{D}})\,{}_1F_1[1+\tfrac{r_1}{4},\tfrac32,\tfrac{4}{D}]+\sqrt{D}H_{-1-\tfrac{r_1}{2}}(\tfrac{2}{\sqrt{D}})\,{}_1F_1[\tfrac{r_1}{4},\tfrac12,\tfrac{4}{D}]\right)}\\
        &\times\frac{1}{\beta H_{-\tfrac{r_2}{2}}(\tfrac{\beta}{\sqrt{4D}})\,{}_1F_1[1+\tfrac{r_2}{4},\tfrac32,\tfrac{\beta^2}{4D}]+2\sqrt{D}H_{-1-\tfrac{r_2}{2}}(\tfrac{\beta}{\sqrt{4D}})\,{}_1F_1[\tfrac{r_2}{4},\tfrac12,\tfrac{\beta^2}{4D}]}\\
        &\times \Big(\\
        &H_{-r_1}(\tfrac{x_0-4}{\sqrt{2D}})\Big(2H_{-\tfrac{r_1}{2}}(\tfrac{2}{\sqrt{D}})\,{}_1F_1[1+\tfrac{r_1}{4},\tfrac32,\tfrac{4}{D}]+\sqrt{D}H_{-1-\tfrac{r_1}{2}}(\tfrac{2}{{\sqrt{D}}})\,{}_1F_1[\tfrac{r_1}{4},\tfrac12,\tfrac{4}{D}]\Big)\\
        &\times \Big((r_1-r_2)\beta H_{-\tfrac{r_2}{2}}(\tfrac{L}{\sqrt{D}})\,{}_1F_1[1+\tfrac{r_2}{4},\tfrac32,\tfrac{\beta^2}{4D}]-r_1\beta H_{-\tfrac{r_2}{2}}(\tfrac{\beta}{\sqrt{4D}})\,{}_1F_1[1+\tfrac{r_2}{4},\tfrac32,\tfrac{\beta^2}{4D}]\\
        &+2\sqrt{D}H_{-1-\tfrac{r_2}{2}}(\tfrac{\beta}{\sqrt{4D}})\big((r_1-r_2){}_1F_1[\tfrac{r_2}{4},\tfrac12,\tfrac{L^2}{D}]-r_1{}_1F_1[\tfrac{r_2}{4},\tfrac12,\tfrac{\beta^2}{4D}]\big)\Big)\\
        &+\sqrt{2D}H_{-1-r_1}(-\sqrt{\tfrac{2}{D}})\Big(r_2H_{-\tfrac{r_1}{2}}(\tfrac{\beta}{\sqrt{4D}})\,{}_1F_1[\tfrac{r_1}{4},\tfrac12,\tfrac{4}{D}]\Big(\beta H_{-\tfrac{r_2}{2}}(\tfrac{L}{\sqrt{D}})\,{}_1F_1[1+\tfrac{r_2}{4},\tfrac32,\tfrac{\beta^2}{4D}]+2\sqrt{D}H_{-1-\tfrac{r_2}{2}}(\tfrac{\beta}{\sqrt{4D}})\,{}_1F_1[\tfrac{r_2}{4},\tfrac12,\tfrac{L^2}{D}]\Big)\\
        &+2r_1\sqrt{D}H_{-1-\tfrac{r_1}{2}}(\tfrac{\beta}{\sqrt{4D}})\,{}_1F_1[\tfrac{r_1}{4},\tfrac12,\tfrac{4}{D}]\Big(H_{-\tfrac{r_2}{2}}(\tfrac{L}{\sqrt{D}})\,{}_1F_1[\tfrac{r_2}{4},\tfrac12,\tfrac{\beta^2}{4D}]-H_{-\tfrac{r_2}{2}}(\tfrac{\beta}{\sqrt{4D}})\,{}_1F_1[\tfrac{r_2}{4},\tfrac12,\tfrac{L^2}{D}]\Big)\\
        &-H_{-\tfrac{r_1}{2}}(\tfrac{2}{\sqrt{D}})\Big({}_1F_1[\tfrac{r_2}{4},\tfrac12,\tfrac{L^2}{D}]\big(r_1\beta H_{-\tfrac{r_2}{2}}(\tfrac{\beta}{\sqrt{4D}})\,{}_1F_1[1+\tfrac{r_1}{4},\tfrac32,\tfrac{\beta^2}{4D}]+2r_2\sqrt{D}H_{-1-\tfrac{r_2}{2}}(\tfrac{\beta}{\sqrt{4D}})\,{}_1F_1[\tfrac{r_1}{4},\tfrac12,\tfrac{\beta^2}{4D}]\big)\\
        &+\beta H_{-\tfrac{r_2}{2}}(\tfrac{L}{\sqrt{D}})\big(r_2{}_1F_1[\tfrac{r_1}{4},\tfrac12,\tfrac{\beta^2}{4D}]{}_1F_1[1+\tfrac{r_2}{4},\tfrac32,\tfrac{\beta^2}{4D}]-r_1{}_1F_1[1+\tfrac{r_1}{4},\tfrac32,\tfrac{\beta^2}{4D}]{}_1F_1[\tfrac{r_2}{4},\tfrac12,\tfrac{\beta^2}{4D}]\big)\Big)\Big)\\
        &+H_{-r_1}(-\sqrt{\tfrac{2}{D}})\Big(2r_2 H_{-\tfrac{r_1}{2}}(\tfrac{\beta}{\sqrt{4D}})\,{}_1F_1[1+\tfrac{r_1}{4},\tfrac32,\tfrac{4}{D}]\Big(\beta H_{-\tfrac{r_2}{2}}(\tfrac{L}{\sqrt{D}})\,{}_1F_1[1+\tfrac{r_2}{4},\tfrac32,\tfrac{\beta^2}{4D}]+2\sqrt{D}H_{-1-\tfrac{r_2}{2}}(\tfrac{\beta}{\sqrt{4D}})\,{}_1F_1[\tfrac{r_2}{4},\tfrac12,\tfrac{L^2}{D}]\Big)\\
        &+\sqrt{D}\Big(4r_1H_{-1-\tfrac{r_1}{2}}(\tfrac{\beta}{\sqrt{4D}})\,{}_1F_1[1+\tfrac{r_1}{4},\tfrac32,\tfrac{4}{D}]\Big(H_{-\tfrac{r_2}{2}}(\tfrac{L}{\sqrt{D}})\,{}_1F_1[\tfrac{r_2}{4},\tfrac{1}{2},\tfrac{\beta^2}{4D}]-H_{-\tfrac{r_2}{2}}(\tfrac{\beta}{\sqrt{4D}})\,{}_1F_1[\tfrac{r_2}{4},\tfrac12,\tfrac{L^2}{D}]\\
        &+H_{-1-\tfrac{r_1}{2}}(\tfrac{2}{\sqrt{D}})\Big({}_1F_1[\tfrac{r_2}{4},\tfrac12,\tfrac{L^2}{D}]\Big(r_1\beta H_{-\tfrac{r_2}{2}}(\tfrac{\beta}{\sqrt{4D}})\,{}_1F_1[1+\tfrac{r_1}{4},\tfrac32,\tfrac{\beta^2}{4D}]+2r_2\sqrt{D}H_{-1-\tfrac{r_2}{2}}(\tfrac{\beta}{\sqrt{4D}})\,{}_1F_1[\tfrac{r_1}{4},\tfrac12,\tfrac{\beta^2}{4D}]\Big)\\
        &+\beta H_{-\tfrac{r_2}{2}}(\tfrac{L}{\sqrt
        D})\Big(r_2{}_1F_1[\tfrac{r_1}{4},\tfrac12,\tfrac{\beta^2}{4D}]{}_1F_1[1+\tfrac{r_2}{4},\tfrac32,\tfrac{\beta^2}{4D}]-r_1{}_1F_1[1+\tfrac{r_1}{4},\tfrac32,\tfrac{\beta^2}{4D}]{}_1F_1[\tfrac{r_2}{4},\tfrac12,\tfrac{\beta^2}{4D}]\Big)\Big)\Big)\Big)\Big)\\
        &\Big).
    \end{aligned}
\end{equation}

\section{The MFPT for piecewise resetting in two dimensions, with smooth quadratic potential}\label{app:MFPT2D}
The MFPT for resetting in a smooth two-dimensional potential, $V(R)=R^2$, with a piecewise constant resetting rate given by Eq.\eqref{eq:resetCon_multidim}, is given by
\begin{equation}\label{eq:MFPT2D}\footnotesize
\begin{aligned}
    T_0=&\frac{1}{r_1 r_2 \left(e^{\frac{i \pi  r_1}{4}} \Gamma \left(\frac{r_1}{4}\right) \, _1F_1\left(\frac{r_1}{4};1;\frac{R_0^2}{D
   }\right)+G_{1,2}^{2,0}\left(-\frac{R_0^2}{D }|
\begin{array}{c}
 1-\frac{r_1}{4} \\
 0,0 \\
\end{array}
\right)\right)}\\&\times \frac{1}{\left(r_1 \, _1F_1\left(\frac{r_1}{4}+1;2;\frac{\beta ^2}{4 D }\right) G_{1,2}^{2,0}\left(-\frac{\beta ^2}{4 D }|
\begin{array}{c}
 1-\frac{r_1}{4} \\
 0,0 \\
\end{array}
\right)-4 \, _1F_1\left(\frac{r_1}{4};1;\frac{\beta ^2}{4 D }\right) G_{1,2}^{2,0}\left(-\frac{\beta ^2}{4 D }|
\begin{array}{c}
 -\frac{r_1}{4} \\
 -1,0 \\
\end{array}
\right)\right)}\\
    &\times\Big(r_2 \Big(e^{\frac{i \pi  r_1}{4}} r_1 \Gamma \left(\frac{r_1}{4}\right) \, _1F_1\left(\frac{r_1}{4};1;\frac{R_0^2}{D }\right)
   \, _1F_1\left(\frac{r_1}{4}+1;2;\frac{\beta ^2}{4 D }\right)-e^{\frac{i \pi  r_1}{4}} r_1 \Gamma \left(\frac{r_1}{4}\right) \,
   _1F_1\left(\frac{r_1}{4};1;\frac{R_0^2}{D }\right) \, _1F_1\left(\frac{r_1}{4}+1;2;\frac{\beta ^2}{4 D }\right)\\
   &+e^{\frac{i \pi  r_1}{4}}
   r_1 \Gamma \left(\frac{r_1}{4}\right) \, _1F_1\left(\frac{r_1}{4};1;\frac{\beta ^2}{4 D }\right) \, _1F_1\left(\frac{r_1}{4}+1;2;\frac{\beta
   ^2}{4 D }\right)+(r_1-r_1) G_{1,2}^{2,0}\left(-\frac{R_0^2}{D }|
\begin{array}{c}
 1-\frac{r_1}{4} \\
 0,0 \\
\end{array}
\right) \, _1F_1\left(\frac{r_1}{4}+1;2;\frac{\beta ^2}{4 D }\right)\\
&+r_1 G_{1,2}^{2,0}\left(-\frac{\beta ^2}{4 D }|
\begin{array}{c}
 1-\frac{r_1}{4} \\
 0,0 \\
\end{array}
\right) \, _1F_1\left(\frac{r_1}{4}+1;2;\frac{\beta ^2}{4 D }\right)-e^{\frac{i \pi  r_1}{4}} r_1 \Gamma \left(\frac{r_1}{4}\right) \,
   _1F_1\left(\frac{r_1}{4}+1;2;\frac{\beta ^2}{4 D }\right) \, _1F_1\left(\frac{r_1}{4};1;\frac{\beta ^2}{4 D }\right)\\
   &-4 \,
   _1F_1\left(\frac{r_1}{4};1;\frac{\beta ^2}{4 D }\right) G_{1,2}^{2,0}\left(-\frac{\beta ^2}{4 D }|
\begin{array}{c}
 -\frac{r_1}{4} \\
 -1,0 \\
\end{array}
\right)\Big) G_{1,2}^{2,0}\left(-\frac{L^2}{D }|
\begin{array}{c}
 1-\frac{r_1}{4} \\
 0,0 \\
\end{array}
\right)\\
&-r_1 \Big(e^{\frac{i \pi  r_1}{4}} r_1 \Gamma \left(\frac{r_1}{4}\right) \left(\, _1F_1\left(\frac{r_1}{4};1;\frac{R_0^2}{D
   }\right) \, _1F_1\left(\frac{r_1}{4}+1;2;\frac{\beta ^2}{4 D }\right)-\, _1F_1\left(\frac{r_1}{4}+1;2;\frac{\beta ^2}{4 D }\right) \,
   _1F_1\left(\frac{r_1}{4};1;\frac{L^2}{D }\right)\right)\\
   &+r_1 \, _1F_1\left(\frac{r_1}{4}+1;2;\frac{\beta ^2}{4 D }\right)
   G_{1,2}^{2,0}\left(-\frac{R_0^2}{D }|
\begin{array}{c}
 1-\frac{r_1}{4} \\
 0,0 \\
\end{array}
\right)-4 \, _1F_1\left(\frac{r_1}{4};1;\frac{L^2}{D }\right) G_{1,2}^{2,0}\left(-\frac{\beta ^2}{4 D }|
\begin{array}{c}
 -\frac{r_1}{4} \\
 -1,0 \\
\end{array}
\right)\Big) G_{1,2}^{2,0}\left(-\frac{\beta ^2}{4 D }|
\begin{array}{c}
 1-\frac{r_1}{4} \\
 0,0 \\
\end{array}
\right)\\
&-4 \Big(e^{\frac{i \pi  r_1}{4}} \Gamma \left(\frac{r_1}{4}\right) \left(r_1 \, _1F_1\left(\frac{r_1}{4};1;\frac{\beta ^2}{4 D }\right)
   \, _1F_1\left(\frac{r_1}{4};1;\frac{L^2}{D }\right)+\, _1F_1\left(\frac{r_1}{4};1;\frac{R_0^2}{D }\right) \left((r_1-r_1) \,
   _1F_1\left(\frac{r_1}{4};1;\frac{L^2}{D }\right)-r_1 \, _1F_1\left(\frac{r_1}{4};1;\frac{\beta ^2}{4 D
   }\right)\right)\right)\\
   &+\left((r_1-r_1) \, _1F_1\left(\frac{r_1}{4};1;\frac{L^2}{D }\right)-r_1 \, _1F_1\left(\frac{r_1}{4};1;\frac{\beta
   ^2}{4 D }\right)\right) G_{1,2}^{2,0}\left(-\frac{R_0^2}{D }|
\begin{array}{c}
 1-\frac{r_1}{4} \\
 0,0 \\
\end{array}
\right)\\
&+r_1 \, _1F_1\left(\frac{r_1}{4};1;\frac{L^2}{D }\right) G_{1,2}^{2,0}\left(-\frac{\beta ^2}{4 D }|
\begin{array}{c}
 1-\frac{r_1}{4} \\
 0,0 \\
\end{array}
\right)\Big) G_{1,2}^{2,0}\left(-\frac{\beta ^2}{4 D }|
\begin{array}{c}
 -\frac{r_1}{4} \\
 -1,0 \\
\end{array}
\right)\Big)
\end{aligned}
\end{equation}
Here
\begin{equation*}
G^{m,n}_{p,q}\left(z|\begin{array}{c}
 a_1,\dots, a_p \\
 b_1,\dots,b_q \\
\end{array}\right)
\end{equation*}
is the Meijer $G$-function. 

\section{Additional results on the MFPT when $r_2>r_1$}
\begin{figure}[tp]
\centering
    \begin{subfigure}{.48\linewidth}
    \includegraphics[width=\linewidth]{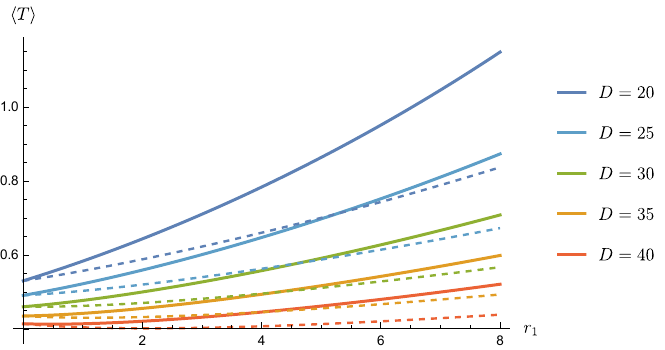}
    \caption{$r_2 = 10 r_1$}
    \label{subfig:compare_smooth3_a}
    \end{subfigure}
    \hfill
    \begin{subfigure}{.48\linewidth}
    \includegraphics[width=\linewidth]{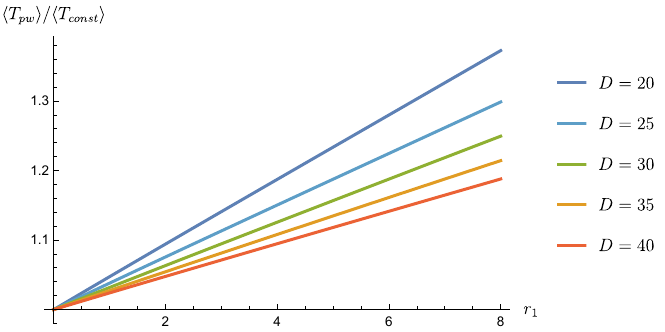}
    \caption{$r_2 = 10 r_1$}
    \label{subfig:compare_smooth3_b}
    \end{subfigure}
\caption{(1d smooth potential / piecewise vs. constant reset / varying $D$) Comparison of the MFPT, $\langle{T}\rangle \coloneqq T_0$ in Eq.~\eqref{eq:smooth_sol_2r}, between using a constant reset rate vs. the piecewise-constant reset rate defined in Eq.~\eqref{eq:2resets} with $r_2 = 10 r_1$, for the smooth quadratic potential $V(x) = x^2$, for values of $D \in \{20, 25, 30, 35, 40\}$. Here we have fixed $L=0.01$, $x_0=4$, $\beta=1$.
\newline
(a) MFPT for constant reset rate ($\langle{T_{const}}\rangle$, dashed) and piecewise-constant reset rate ($\langle{T_{pw}}\rangle$, solid), labelled by $D$.
(b) Ratio $\langle{T_{pw}}\rangle / \langle{T_{const}}\rangle$, labelled by $D$.
All numerical evaluations were done using Mathematica \cite{Mathematica}.
}
\label{fig:compare_smooth3}
    \end{figure}

\begin{figure}[tp]
\centering
    \begin{subfigure}{.48\linewidth}
    \includegraphics[width=\linewidth]{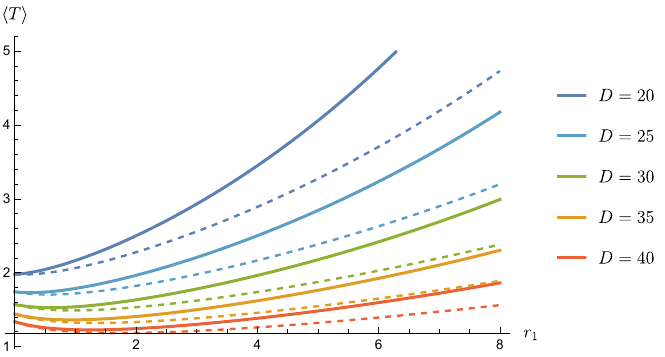}
    \caption{$r_2 = 10 r_1$}
    \label{subfig:compare_nonsmooth3_a}
    \end{subfigure}
    \hfill
    \begin{subfigure}{.48\linewidth}
    \includegraphics[width=\linewidth]{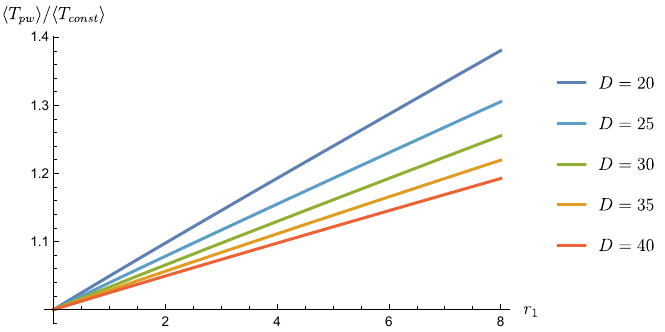}
    \caption{$r_2 = 10 r_1$}
    \label{subfig:compare_nonsmooth3_b}
    \end{subfigure}
\caption{(1d non-smooth potential / piecewise vs. constant reset / varying $D$) 
Comparison of the MFPT, $\langle{T}\rangle \coloneqq T_0$ in Eq.~\eqref{eq:nonsmooth_sol_2r}, between using a constant reset rate and the piecewise-constant reset rate defined in Eq.~\eqref{eq:2resets} with $r_2 = 10 r_1$, for the non-smooth piecewise-quadratic potential defined in Eq.~\eqref{eq:piecewisepot_1d}, for values of $D \in \{20, 25, 30, 35, 40\}$. 
Here we have fixed $L=0.01$, $x_0=6$, $\beta=1$. \newline
(a) MFPT for constant reset rate ($\langle{T_{const}}\rangle$, dashed) and piecewise-constant reset rate ($\langle{T_{pw}}\rangle$, solid), labelled by $D$. (b) Ratio $\langle{T_{pw}}\rangle / \langle{T_{const}}\rangle$, labelled by $D$.
All numerical evaluations were done using Mathematica \cite{Mathematica}.
}
\label{fig:compare_nonsmooth3}
    \end{figure}

\begin{figure}[tp]
\centering
    \begin{subfigure}{.48\linewidth}
    \includegraphics[width=\linewidth]{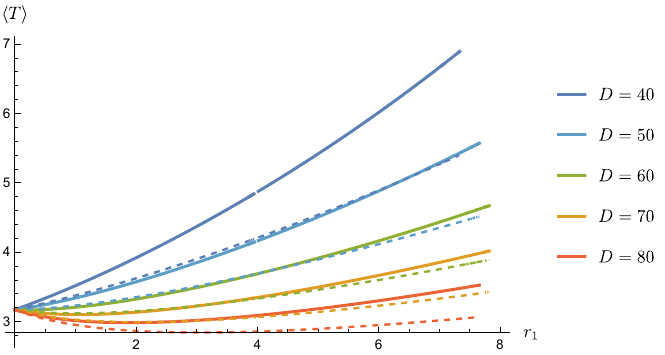}
    \caption{$r_2 = 10 r_1$}
    \label{subfig:compare_nonsmooth23d_a}
    \end{subfigure}
    \hfill
    \begin{subfigure}{.48\linewidth}
    \includegraphics[width=\linewidth]{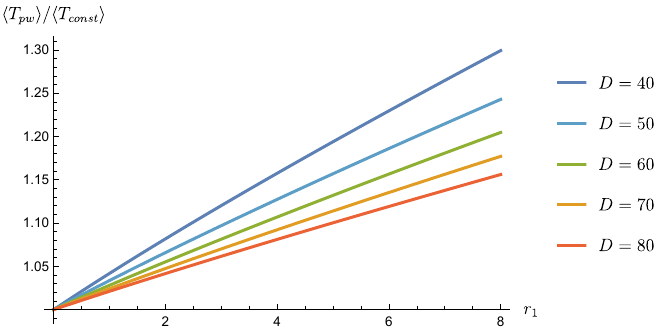}
    \caption{$r_2 = 10 r_1$}
    \label{subfig:compare_nonsmooth23d_b}
    \end{subfigure}
\caption{(2d smooth potential / piecewise vs. constant reset / varying $D$) 
Comparison of the MFPT, $\langle{T}\rangle \eqqcolon T_0$ of Eq. \eqref{eq:MFPT2D}, between using a constant reset rate and the piecewise-constant reset rate defined in Eq.~\eqref{eq:resetCon_multidim} with $r_2 = 10 r_1$, for a 2-dimensional smooth quadratic potential $V(R)=R^2$, for values of $D \in \{40, 50, 60, 70, 80\}$.
Here we have fixed $L=0.01$, $R_0=4\sqrt{2}$, $\beta=1$. \newline
(a) MFPT for constant reset rate ($\langle{T_{const}}\rangle$, dashed) and piecewise-constant reset rate ($\langle{T_{pw}}\rangle$, solid), labelled by $D$. (b) Ratio $\langle{T_{pw}}\rangle / \langle{T_{const}}\rangle$, labelled by $D$.
All numerical evaluations were done using Mathematica \cite{Mathematica}.
}
\label{fig:compare_2D_r2_10r1}
    \end{figure}

For completeness of the discussion, we add here a few more figures, Figs. \ref{fig:compare_smooth3}-\ref{fig:compare_2D_r2_10r1}, which correspond to Figs. \ref{fig:compare_smooth2}, \ref{fig:compare_nonsmooth2} and \ref{fig:compare_2D_r2_01r1}, but with $r_2=10r_1$ instead. This highlights the fact that it is only when $r_2<r_1$ that we see the benefit of having multiple reset rates.

\section{Example trajectory of the discrete-time 2d diffusion process}
\label{app:trajectory_2d}

\begin{figure}
    \centering
    \includegraphics[width=0.8\linewidth]{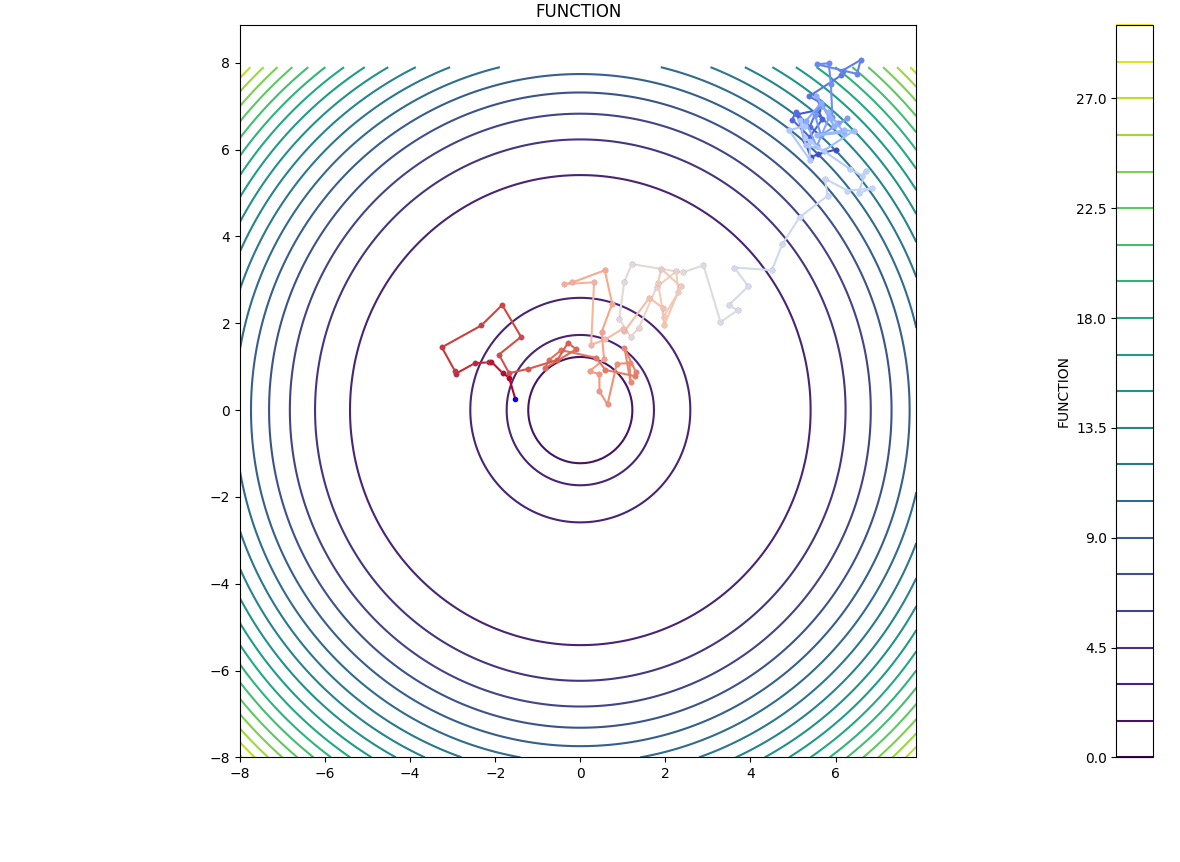}
    \caption{Example of a particle diffusing on a 2d non-smooth piecewise quadratic potential having a gradient discontinuity at a radius $R = 2$, overlaid on the potential's contour plot. The diffusion and potential characteristics are described in App.~\ref{app:trajectory_2d}. Time is represented by the trajectory color, as it turns from blue to red.
    }
    \label{fig:trajectory_2d}
\end{figure}

Fig.~\ref{fig:trajectory_2d} shows an example of a trajectory of a particle driven by the discrete-time diffusion equation \eqref{eq:diffusion_discrete} with $F(x) = -\nabla V(x)$, where $V(x)$ is a 2d symmetric non-smooth piecewise quadratic potential defined by Eq.~\eqref{eq:piecewisepot_1d} in each radial direction. A piecewise-constant reset rate strategy is used, as defined in Eq.~\eqref{eq:resetCon_multidim}, $\beta = 1$, $r_1 = 10$, and $r_2 = 10r_1$.
The reset point is set to $x_0 = (6, 6)$, the diffusion coefficient is $D = 810$. The discretization interval of the original continuous-time diffusion equation \eqref{eq:diffusion_equation} is set to $\Delta t = 10^{-4}$.
The particle initially tends to drift away but finally gets closer to the potential's global optimum at $(0, 0)$.

\bibliographystyle{unsrt}
\normalem
\bibliography{references}
\pagebreak

\end{document}